\documentclass[a4paper,11pt]{article}
\pdfoutput=1
\usepackage{jheppub}
\usepackage{color}
\usepackage{array}
\newcolumntype{C}[1]{>{\centering\arraybackslash}p{#1}}
\usepackage{slashed,dsfont}
\usepackage{bm,wasysym}
\usepackage{longtable}
\usepackage{booktabs}

\usepackage{tabularx}
\usepackage{rotating}
\usepackage{lscape} 
\usepackage[normalem]{ulem}

\newcommand{\lsim}{
\mathrel{\hbox{\rlap{\hbox{\lower4pt\hbox{$\sim$}}}\hbox{$<$}}}}

\newcommand{\gsim}{
\mathrel{\hbox{\rlap{\hbox{\lower4pt\hbox{$\sim$}}}\hbox{$>$}}}}

\renewcommand{\arraystretch}{2}

\allowdisplaybreaks[2]

\newcommand{\nn}{\nonumber}

\def\Lb{{\Lambda_b}}
\def\Lst{{\Lambda^\ast}}
\def\mLb{{m_{\Lambda_b}}}
\def\mmLb{{m^2_{\Lambda_b}}}
\def\mL{{m_{\Lambda^\ast}}}
\def\mmL{{m^2_{\Lambda^\ast}}}
\def\plpl{{+\frac{1}{2}+\frac{1}{2}}}
\def\plmi{{+\frac{1}{2}-\frac{1}{2}}}
\def\mipl{{-\frac{1}{2}+\frac{1}{2}}}
\def\mimi{{-\frac{1}{2}-\frac{1}{2}}}
\def\re{{\rm Re}}  \def\im{{\rm Im}}
\def\mC{{\mathcal{C}}}

\def\cl{{\cos\theta_\ell}}
\def\ccl{{\cos^2\theta_\ell}}
\def\sl{{\sin\theta_\ell}}
\def\ssl{{\sin^2\theta_\ell}}
\def\cLst{{\cos\theta_{\Lst}}}
\def\ccLst{{\cos^2\theta_{\Lst}}}
\def\sLst{{\sin\theta_{\Lst}}}
\def\ssLst{{\sin^2\theta_{\Lst}}}

\definecolor{schrift}{RGB}{120,0,0}

\def\apasl{{A_{\parallel \rm S}^L}}
\def\apasr{{A_{\parallel \rm S}^R}}  
\def\apaol{{A_{\parallel 0}^L}} 
\def\apaor{{A_{\parallel 0}^R}}
\def\apanl{{A_{\parallel 1}^L}} 
\def\apanr{{A_{\parallel 1}^R}}
\def\bpanl{{B_{\parallel 1}^L}}
\def\bpanr{{B_{\parallel 1}^R}}
\def\apatl{{A_{\parallel t}^L}}
\def\apatr{{A_{\parallel t}^R}}

\def\apesl{{A_{\perp \rm S}^L}}
\def\apesr{{A_{\perp \rm S}^R}}
\def\apeol{{A_{\perp 0}^L}}
\def\apeor{{A_{\perp 0}^R}}
\def\apenl{{A_{\perp 1}^L}} 
\def\apenr{{A_{\perp 1}^R}} 
\def\bpenl{{B_{\perp 1}^L}} 
\def\bpenr{{B_{\perp 1}^R}}
\def\apetl{{A_{\perp t}^L}}
\def\apetr{{A_{\perp t}^R}}

\title{\boldmath\color{schrift}{The $\Lambda_b\to\Lambda^\ast(1520)(\to N\!\bar{K})\ell^+\ell^-$ decay at low-recoil in HQET}}

\author{Diganta Das, Jaydeb Das}
\affiliation{{\sf Department of Physics and Astrophysics, University of Delhi, Delhi 110007, India}}
\emailAdd{diganta99@gmail.com}
\emailAdd{jaydebphysics@gmail.com}

\abstract{In this paper we discuss the Standard Model and new physics sensitivity of the $\Lambda_b\to\Lambda^\ast(1520)(\to N\!\bar{K})\ell^+\ell^-$ decay at low-recoil, where $\ell^\pm$ are massive leptons and $N\!\bar{K}=\{pK^-, n\bar{K}^0 \}$. We provide a full angular distribution with a set of operators that includes the Standard Model operators and their chirality flipped counterparts, and new scalar and pseudo-scalar operators. The resulting angular distribution allows us to construct observables that we study in the Standard Model and in model-independent new physics scenarios. To reduce the hadronic effects emanating from the $\Lambda_b\to\Lambda^\ast(1520)$ transition form factors, we exploit the Heavy Quark Effective theory framework valid at low $\Lambda^\ast(1520)$ recoil, \emph{i.e.,} large dilepton invariant mass squared $q^2\sim \mathcal{O}(m_b^2)$. Working to the leading order in $1/m_b$ and including $\mathcal{O}(\alpha_s)$ corrections, we compute the `improved Isgur-Wise relations' between the form factors. The relations correlate the form factors and thereby allow the description of this decay at the low-recoil region with a smaller number of independent form factors.} 


\begin{document}

\maketitle

\renewcommand{\arraystretch}{1.6}

\section{Introduction}
A large number of experimental measurements in $b\to s\ell^+\ell^-$ transitions have shown deviations from the Standard Model (SM) predictions. The most recent of these deviations are the ratio of $B\to K^{(\ast)}\ell^+\ell^-$ branching ratios for muon over electrons in the measurements of $R_K$ \cite{Aaij:2019wad} and $R_{K^\ast}$ \cite{Aaij:2017vbb} observables which could be hints of lepton flavor universality (LFU) violation. The other deviations include the branching ratios of $B\to K\mu^+\mu^-$ \cite{Aaij:2014pli}, $B\to K^\ast\mu^+\mu^-$ \cite{Aaij:2013iag, Aaij:2016flj}, $B_s\to\phi\mu^+\mu^-$ \cite{Aaij:2015esa}, and the `optimized' observables in the $B\to K^\ast\mu^+\mu^-$ decay \cite{Aaij:2020nrf}. These deviations have put the exploration of new physics (NP) in the $b\to s\ell^+\ell^-$ transitions at the forefront of the $B$-physics program. 

Indeed, the LHCb's recent capability to study baryonic decays has opened up a new avenue to explore the $b\to s\ell^+\ell^-$ transitions. For instance, the $\Lb\to\Lambda(1116)(\to N\pi)\mu^+\mu^-$ has been measured by the LHCb \cite{Aaij:2018gwm, Aaij:2015xza}, and the observed branching ratio is lower than the SM predictions \cite{Gutsche:2013pp, Boer:2014kda, Das:2018sms, Blake:2017une} with a trend that is also observed in the $B_{(s)}\to K^{(\ast)}(\phi)\mu^+\mu^-$ decays. The LHCb has also performed the LFU violation measurement in $\Lb\to pK^-\ell^+\ell^-$ decay \cite{Aaij:2019bzx}.

Theoretical studies of $\Lb$ decay through a $b\to s\ell^+\ell^-$ transition have received considerable attention in the recent years \cite{Gutsche:2013pp, Boer:2014kda, Roy:2017dum, Das:2018sms, Das:2018iap, Blake:2017une, Descotes-Genon:2019dbw, Yan:2019tgn}. Among the different semileptonic modes of $\Lb$ decays to hadrons, the decay to $\Lst\equiv \Lst(1520)$ has the dominant contribution \cite{Aaij:2015tga}. The $\Lst$ has spin parity of $J^P=3/2^-$ and decays strongly to the $N\!\bar{K}$ pair. These characteristics make the $\Lst$ easily distinguishable from the closely lying $\Lambda(1600)$, $\Lambda(1405)$, and the weakly decaying $\Lambda(1116)$, all of which have spin parity $J^P=1/2^\pm$. 

In this paper we discuss some aspects of $\Lambda_b\to\Lst(\to N\!\bar{K})\ell^+\ell^-$ decay in the SM and beyond. The four-body decay proceeds through the weak decay of $\Lb\to\Lst\ell^+\ell^-$ and the subsequent strong decay $\Lst\to N\!\bar{K}$. The SM operators for the weak decay is supplemented with their chirality flipped counterparts (henceforth called SM$^\prime$ operators), and a scalar and pseudo-scalar operators (henceforth SP operators). We provide a full angular distribution and express the angular coefficients in terms of transversity amplitudes. The masses of the leptons in the final state are retained which can be important if the dilepton pair is heavy. From the four-fold angular distribution, we construct several observables that we study in the SM and in model-independent NP. 

Like any exclusive decay, this mode also suffers from long-distance QCD dynamics coming from different sources. This includes the ``naively" factorizable contributions of the $b\to s\ell^+\ell^-$ and $b\to s\gamma$ operators parametrized in terms of form factors. In two opposite kinematical regions, significant reduction to the number of independent form factors can be achieved by means of effective theories--Soft-Collinear Effective Theory (SCET) \cite{Bauer:2000yr, Beneke:2000wa, Beneke:2001at, Beneke:2004dp, Charles:1998dr} at large hadronic recoil or low dilepton invariant mass squared $q^2$, and the Heavy Quark Effective Theory (HQET) \cite{Isgur:1989ed, Isgur:1990pm, Isgur:1989vq} at low-recoil or large $q^2$. A systematic discussion of the decay in the SCET framework \cite{Mannel:2011xg,Wang:2011uv, Feldmann:2011xf} is beyond the scope of the present paper due to poor knowledge of the baryonic wave function and the fact that the spectator scattering effects are more complicated to deal with \cite{Wang:2015ndk}.  We will rather focus on the low-recoil region where the decay can be described by the HQET framework, and an operator product (OPE) expansion in $1/Q$ \cite{Grinstein:2004vb}, where $Q\sim (m_b, \sqrt{q^2})$. The low-recoil OPE allows us to include the contributions from the charm quark loops into the Wilson coefficients \cite{Grinstein:2004vb}. The HQET spin flavor symmetry ensures relations between the form factors, known as the Isgur-Wise relations \cite{Isgur:1990kf}. Following the prescription of Ref.~\cite{Grinstein:2004vb}, we obtain `improved Isgur-Wise relations' to leading order in $1/m_b$ and including $\mathcal{O}(\alpha_s)$ corrections. By means of these relations, the form factors are correlated and each of the transversity amplitudes depends on a single form factor, as a result of which the short- and long-distance physics factorize in the angular coefficients. The reduction in the number of independent form factors leads to improved predictions in the low-recoil region. The low-recoil factorization allows us to construct observables that are sensitive to NP and have reduced dependence on form factor inputs.

The article is organized as follows. In the Sec.~\ref{sec:kin} we describe the effective Hamiltonian for the $b\to s\ell^+\ell^-$ transition in the SM+SM$^\prime$+SP set of operators and derive the decay amplitudes. The four-fold angular distribution is worked out and the angular coefficients are derived in the Sec.~\ref{sec:angular}. In Sec.~\ref{sec:obs} we construct the observables. In Sec.~\ref{sec:fffree} we derive the improved Isgur-Wise relations between the $\Lb\to\Lst$ form factors, describe the low-recoil factorization, and construct observables for clean extraction of the electroweak physics. We present our numerical analysis in Sec.~\ref{sec:num} and a summary is given in Sec.~\ref{sec:summary}. We have also presented the necessary formulas in the appendices.

\section{The Framework \label{sec:kin}}
\subsection{Effective Hamiltonian \label{subsec:effHam}}
The $\Lb\to\Lst\ell^+\ell^-$ decay is governed by the $b\to s\ell^+\ell^-$ transition for which we assume the following effective Hamiltonian
\begin{equation}\label{eq:Heff1}
\mathcal{H}^{\rm eff} = - \frac{4G_F}{\sqrt{2}}V_{tb}V_{ts}^\ast\frac{\alpha_e}{4\pi}  \bigg(\sum_i \mC_i \mathcal{O}_i + \sum_j  \mC_{j^\prime} \mathcal{O}_{j^\prime} \bigg)\, ,
\end{equation}
where $i = 7, 9, 10, S, P$ and $j^\prime = 7^\prime, 9^\prime, 10^\prime, S^\prime, P^\prime$. The operators read 
\begin{eqnarray}\label{eq:opbasis}
\begin{split}
&O_7 = \frac{m_b}{e} \big[\bar{s}\sigma^{\mu\nu}P_{R}b\big]F_{\mu\nu}\, ,\quad O_{7^\prime} = \frac{m_b}{e} \big[\bar{s}\sigma^{\mu\nu}P_{L}b\big]F_{\mu\nu}\, ,\\ & \mathcal{O}_9 = \big[\bar{s}\gamma^\mu P_{L}b \big]\big[\ell\gamma_\mu\ell \big]\, ,
\quad \mathcal{O}_{9^{\prime}} = \big[\bar{s}\gamma^\mu P_{R}b \big]\big[\ell\gamma_\mu\ell \big]\, ,\\
&\mathcal{O}_{10} = \big[\bar{s}\gamma^\mu P_{L}b \big]\big[\ell\gamma_\mu\gamma_5\ell \big]\, , \quad
\mathcal{O}_{10^{\prime}} =  \big[\bar{s}\gamma^\mu P_{R}b \big]\big[\ell\gamma_\mu\gamma_5\ell \big]\, ,\\
&\mathcal{O}_{S^{(\prime)}} = \big[\bar{s}P_{R(L)}b \big]\big[\ell\ell \big]\, ,\quad \mathcal{O}_{P^{(\prime)}} = \big[\bar{s}P_{R(L)}b \big]\big[\ell\gamma_5\ell \big]\, .
\end{split}
\end{eqnarray}
In the SM only $\mathcal{O}_{7,9,10}$ appear and the rest of the NP operators may appear in beyond the SM scenarios. NP operators of the types $\mathcal{O}_{9,10}$ may also appear but the effects can be included trivially by the substitutions in the SM Wilson coefficients $\mC_{9,10}\to \mC_{9,10} + \delta \mC_{9,10}^{\rm NP}$. Tensor operators has been ignored in this paper for simplicity. The rest of the parameters are as follows: $G_F$ is the Fermi-constant, $V_{tb}V_{ts}^\ast$ are the Cabibbo-Kobayashi-Maskawa (CKM) elements, $\alpha_e=e^2/4\pi$ is the fine structure constant, and $P_{L(R)}=(1\mp\gamma_5)/2$ are the chirality projectors. The $b$-quark mass appearing in the $\mathcal{O}_{7,7^\prime}$ operators are taken as the running mass in the modified minimal subtraction scheme ($\overline{\rm MS}$). The contribution proportional to $V_{ub}V_{us}^\ast$ has been neglected since $V_{ub}V_{us}^\ast<<V_{tb}V_{ts}^\ast$ and therefore, CP violation is absent.

\subsection{Decay kinematics}
We assign the following momenta and spin variables to the different particles in the decay process 
\begin{eqnarray}
\begin{split}
&\Lambda_b(p,s_\Lb ) \to \Lambda^\ast(k,s_\Lst) \ell^+(q_1) \ell^-(q_2)\, ,\\
&\Lambda^\ast(k,s_\Lst) \to N(k_1,s_N) \bar{K}(k_2)\, ,
\end{split}
\end{eqnarray}
\emph{i.e.,} $p,k,k_1,k_2,q_1$ and $q_2$ are the momenta of $\Lambda_b$, $\Lst$, $N$, $\bar{K}$, and the positively and negatively charged leptons, respectively, and $s_{\Lb,\Lst,N}$ are the projections of the baryon spins on to the $z$-axis in their respective rest frames. For future convenience, we define the momentum for the dilepton pair
\begin{equation}
q^\mu = q_1^\mu + q_2^\mu\, .
\end{equation} 
The momentum conservation gives $k^\mu=k_1^\mu + k_2^\mu$, $p^\mu = k^\mu + q^\mu$. The angle $\theta_\ell$ is defined as the one made by the lepton $\ell^-$ with respect to the $+z$ axis in the $\ell^+\ell^-$ rest frame, $\theta_\Lst$ is the angle made by the nucleon with the $+z$ axis in the $N\!\bar{K}$ rest frame, and $\phi$ is the angle between the decay planes of the dilepton pair and the hadron pair. We have spelled out the kinematics in appendix~\ref{app:kin}.
\subsection{The $\Lambda_b\to\Lambda^\ast\ell^+\ell^-$ decay}
Assuming factorization between the hadronic and the leptonic parts, the matrix element of the four-body decay $\Lb\to\Lambda^\ast(\to N\!\bar{K})\ell^+\ell^-$ can be written as 
\begin{equation}\label{eq:fullM}
\mathcal{M}({s_\Lb,s_N,\lambda_1,\lambda_2}) = \sum_{s_\Lst} \mathcal{M}_{\Lb}^{\lambda_1, \lambda_2}(s_{\Lambda_b},s_
{\Lambda^\star}) \mathcal{M}_{\Lst}(s_\Lst,s_N)\, ,
\end{equation}
where $\mathcal{M}_{\Lst}(s_\Lst,s_N)$ correspond to the matrix element for $\Lst\to N\!\bar{K}$ which is discussed in the next section. The amplitudes for the first stage of the decay can be written as 
\begin{eqnarray}
	\mathcal{M}_{\Lb}^{\lambda_1, \lambda_2}(s_{\Lambda_b},s_
	{\Lambda^\star}) &=& -\frac{4G_F}{\sqrt{2}} V_{tb}V_{ts}^\ast \frac{\alpha_e}{4\pi} \sum_{L(R)} \frac{1}{4} \bigg[\sum_{\lambda} \eta_\lambda H^{L(R)}_{\rm VA,\lambda}(s_\Lb,s_\Lst) L^{\lambda_1,\lambda_2}_{L(R),\lambda} \, \nn\\ &+& H^{L(R)}_{\rm SP}(s_\Lb,s_\Lst) L^{\lambda_1,\lambda_2}_{L(R)} \bigg]\, ,\quad\quad
\end{eqnarray}
where the hadronic and the leptonic helicity amplitudes are defined as the projection of the full hadronic and leptonic amplitudes on to the polarization direction of virtual gauge boson that decays to the dilepton pair, and $\eta_t=1, \eta_{\pm 1, 0}=-1$. Denoting the polarizations of the virtual gauge boson as $\bar{\epsilon}_\mu(\lambda)$ for different polarization states $\lambda=0,\pm 1, t$, the leptonic helicity amplitudes are written as
\begin{align}\label{eq:Ldef1}
& L^{\lambda_1,\lambda_2}_{L(R)} = \langle \bar{\ell}(\lambda_1)\ell(\lambda_2) | \bar{\ell} (1\mp\gamma_5) \ell | 0\rangle\, , \\
& L^{\lambda_1,\lambda_2}_{L(R),\lambda} = \bar{\epsilon}^\mu(\lambda) \langle \bar{\ell}(\lambda_1) \ell(\lambda_2) | \bar{\ell} \gamma_\mu (1\mp\gamma_5) \ell | 0\rangle\, .
\end{align}
Our choice of the gauge boson polarizations and the expressions of $L^{\lambda_1,\lambda_2}_{L(R)}, L^{\lambda_1,\lambda_2}_{L(R),\lambda}$ are detailed in the appendix~\ref{app:pol} and \ref{app:LepHel}, respectively.

The hadronic helicity amplitudes are similarly defined as 
\begin{eqnarray}\label{eq:Hl1}
H^{L(R)}_{{\rm VA},\lambda}(s_\Lb,s_\Lst) &=& \bar{\epsilon}^\ast_\mu(\lambda) \big{\langle}\Lambda(k,s_\Lst)\big| \bigg[ \Big( \mC_9 \mp \mC_{10} \Big)\bar{s} \gamma^\mu (1-\gamma_5)b + (\mC_{9^\prime} \mp \mC_{10^\prime}) \bar{s}\gamma^\mu(1+\gamma_5)b \, \nn\\ &-& \frac{2m_b}{q^2} \bigg(\mC_7 \bar{s}iq_\nu\sigma^{\mu\nu} (1+\gamma_5) b + \mC_{7^\prime} \bar{s}iq_\nu\sigma^{\mu\nu} (1-\gamma_5) b \bigg)\bigg] \big|\Lambda_b(p,s_\Lb)\big\rangle \, ,\\
\label{eq:Hl2}
H^{L(R)}_{\rm SP}(s_\Lb,s_\Lst) &=& \big\langle \Lambda(k,s_\Lst) \big| \bigg[ (\mC_{S^\prime} \mp \mC_{P^\prime}) \bar{s}(1-\gamma_5)b + (\mC_S \mp \mC_P) \bar{s}(1+\gamma_5)b \bigg] \big| \Lambda_b(p,s_\Lb) \big\rangle\, .\nn\\
\end{eqnarray}
In order to calculate the amplitudes, we need to know the form factor parametrizations of the $\Lb\to\Lst$ hadronic matrix elements. We follow the helicity parametrizations \cite{Descotes-Genon:2019dbw} (see appendix~\ref{app:ffParam}) where the $\langle\Lst | \bar s \gamma^\mu b|\Lb\rangle$ ($\langle\Lst | \bar s \gamma^\mu\gamma_5 b|\Lb\rangle$) transition is parametrized in terms of four $q^2$-dependent form factors $f^V_t$, $f^V_0$, $f^V_\perp$, $f^V_g$ ($f^A_t$, $f^A_0$, $f^A_\perp$, $f^A_g$). The matrix elements of the scalar and the pseudo-scalar currents $\langle\Lst | \bar s  b|\Lb\rangle$, $\langle\Lst | \bar s \gamma_5 b|\Lb\rangle$ are obtained from the vector and axial vector matrix elements by the application of equation of motion and depend on $f^V_t$ and $f^A_t$, respectively. The transition $\langle\Lst | \bar si \sigma^{\mu\nu}q_\nu b|\Lb\rangle$ ($\langle\Lst | \bar si \sigma^{\mu\nu}\gamma_5q_\nu b|\Lb\rangle$) is parametrized in terms of three form factors $f^T_0$, $f^T_\perp$, $f^T_g$ ($f^{T5}_0$, $f^{T5}_\perp$, $f^{T5}_g$). Overall, fourteen $q^2$ dependent form factors contribute to this decay.

With the parametrizations of $\Lambda\to\Lambda^\ast$ transitions at our disposal, we now calculate the hadronic amplitudes. In the literatures, instead of the helicity amplitudes, the so-called transversity amplitudes are often used. The transversity amplitudes are linear combinations of the helicity amplitudes, see appendix~\ref{app:TAs}. For the (axial-)vectors currents they read
\begin{align}\label{eq:TAVA1}
& B_{\perp_1}^{L(R)} = \sqrt{2}N \Bigg( f_g^V \sqrt{s_+} \mC_{\rm VA+}^{L(R)} + \frac{2m_b}{q^2} f_g^{T} \sqrt{s_+} (\mC_7 + \mC_7^\prime)  \bigg)\, , \\
& B_{\|_1}^{L(R)} = \sqrt{2}N \Bigg( f_g^A \sqrt{s_-} \mC_{\rm VA-}^{L(R)} + \frac{2m_b}{q^2} f_g^{T5} \sqrt{s_-} (\mC_7 - \mC_7^\prime)  \bigg)\, , \\
& A_{\perp_0}^{L(R)} = -\sqrt{2}N \bigg( f_0^V \frac{(\mLb+\mL)}{\sqrt{q^2}} \frac{s_-\sqrt{s_+}}{\sqrt{6}\mL} \mC_{\rm VA+}^{L(R)} + \frac{2m_b}{q^2} f^T_0 \sqrt{q^2} \frac{s_-\sqrt{s_+}}{\sqrt{6}\mL} (\mC_7 + \mC_7^\prime)\bigg)\, , \\
& A_{\|_0}^{L(R)} = \sqrt{2}N \bigg( f_0^A \frac{(\mLb-\mL)}{\sqrt{q^2}} \frac{s_+\sqrt{s_-}}{\sqrt{6}\mL} \mC_{\rm VA-}^{L(R)} + \frac{2m_b}{q^2} f^{T5}_0 \sqrt{q^2} \frac{s_+\sqrt{s_-}}{\sqrt{6}\mL} (\mC_7 - \mC_7^\prime)\bigg)\, , \\
& A_{\perp_1}^{L(R)} = -\sqrt{2}N \Bigg( f_\perp^V \frac{s_- \sqrt{s_+}}{\sqrt{3}\mL} \mC_{\rm VA+}^{L(R)} + \frac{2m_b}{q^2} f_\perp^T (\mLb+\mL) \frac{s_- \sqrt{s_+}}{\sqrt{3}\mL} (\mC_7 + \mC_7^\prime)  \bigg)\, , \\
& A_{\|_1}^{L(R)} = -\sqrt{2}N \Bigg( f_\perp^A \frac{s_+ \sqrt{s_-}}{\sqrt{3}\mL} \mC_{\rm VA-}^{L(R)} + \frac{2m_b}{q^2} f_\perp^{T5} (\mLb-\mL) \frac{s_+ \sqrt{s_-}}{\sqrt{3}\mL} (\mC_7 - \mC_7^\prime)  \bigg)\, , \\
& A_{\perp_t}^{L(R)} = \mp \sqrt{2}N f^V_t \frac{(\mLb-\mL)}{\sqrt{q^2}} \frac{s_+\sqrt{s_-}}{\sqrt{6}\mL} (\mC_{10} + \mC_{10^\prime} ) \, ,\\
\label{eq:TAVA1end}
& A_{\|_t}^{L(R)} = \pm \sqrt{2}N f^A_t \frac{(\mLb+\mL)}{\sqrt{q^2}} \frac{s_-\sqrt{s_+}}{\sqrt{6}\mL} (\mC_{10} - \mC_{10^\prime} )\, ,
\end{align} 
where $\lambda(a,b,c) = a^2 + b^2 + c^2 - 2(ab+bc+ca)$, and
\begin{equation}
s_\pm = (m_\Lb \pm m_\Lst)^2 - q^2\, .
\end{equation}
The Wilson coefficients $\mC_{\rm V\!A\pm}^{L(R)}$ are defined as follows
\begin{align}
\begin{split}
& \mC_{\rm V\!A+}^{L(R)} = \big( \mC_9 \mp \mC_{10} \big) + \big( \mC_{9^\prime} \mp \mC_{10^\prime} \big)\, ,\\
& \mC_{\rm V\!A-}^{L(R)} = \big( \mC_9 \mp \mC_{10} \big) - \big( \mC_{9^\prime} \mp \mC_{10^\prime} \big)\, .
\end{split}
\end{align}
For the (pseudo-)scalar currents the amplitudes read
\begin{align}
& A_{S\perp}^{L(R)} = \sqrt{2}N f^V_t \frac{\mLb-\mL}{m_b-m_s} \frac{s_+\sqrt{s_-}}{\sqrt{6}\mL} \mC_{\rm SP+}^{L(R)} \, ,\\
& A_{S\|}^{L(R)} = -\sqrt{2}N f^A_t \frac{\mLb+\mL}{m_b+m_s} \frac{s_-\sqrt{s_+}}{\sqrt{6}\mL} \mC_{\rm SP-}^{L(R)}\, ,
\end{align}
where the scalar Wilson coefficients are defined as 
\begin{align}
\begin{split}
&\mC_{\rm SP+}^{L(R)} = (\mC_S + \mC_{S^\prime}) \mp (\mC_P + \mC_{P^\prime})\, ,\\
&\mC_{\rm SP-}^{L(R)} = (\mC_S - \mC_{S^\prime}) \mp (\mC_P - \mC_{P^\prime})\, .
\end{split}
\end{align}
The overall normalization factor that has been customarily absorbed in the transversity amplitude is given by
\begin{equation}
N = G_F V_{tb}V_{ts}^\ast \alpha_e \sqrt{\tau_\Lb \frac{q^2\sqrt{\lambda(\mmLb,\mmL,q^2)}}{3\cdot 2^{11} m_\Lb^3 \pi^5} \beta_\ell \mathcal{B}_\Lst}\, ,\quad \beta_\ell = \sqrt{1-\frac{4m_\ell^2}{q^2}}\, ,
\end{equation}
where $\mathcal{B}_\Lst\equiv\mathcal{B}_\Lst(\Lst\to N\!\bar{K})$ is the branching ratio and $\tau_\Lb$ is the $\Lb$ lifetime.

\subsection{The $\Lambda^\ast\to N\!\bar{K}$ decay  \label{sec:LamStDecay}}
The effective Lagrangian describing the strong decay $\Lambda^\ast\to N\!\bar{K}$ is assumed to be \footnote{A different choice for the Lagrangian is given in \cite{Pascalutsa:1998pw} which leads to same result.} \cite{Nath:1971wp}
\begin{align}\label{eq:Lag2}
&\mathcal{L}_1 = g\mL\bar{\psi}_\mu ( g^{\mu\nu} + a \gamma^\mu \gamma^\nu ) \gamma_5 \Psi \partial_\nu \phi + h.c.,
\end{align}
where $g$ is a coupling constant, $\Psi$ is a spin-1/2 field describing the $\Lb$, and $\phi$ is scalar field corresponding to the $\bar{K}$ meson. The $\Lst$ is a spin 3/2 particle and is described by a Rarita-Schwinger field $\psi_\mu$ \cite{Rarita:1941mf}. In addition to the vector index, there is an implicit spinor index in the Rarita-Schwinger field. The parameter $a$ is only relevant for loop calculations. The Hamiltonian \eqref{eq:Lag2} yields the following matrix element for $\Lst\to N\!\bar{K}$ transition
\begin{align}
&\mathcal{M}^{\Lst}(s_\Lst,s_N) = g \mL k_2^\mu \bar{u}^{s_N} \gamma_5 U^{s_\Lst}_\mu\, ,
\end{align}
where $U^{s_\Lst}_\mu$ is the Rarita-Schwinger spinor describing the $\Lst$ and $u^{s_N}$ is the Dirac spinor for the $N$.  In the rest frame of the $\Lambda^\ast$ the solutions of Rarita-Schwinger and Dirac spinors are given in appendix \ref{app:Rarita}. As can be understood from \eqref{eq:fullM}, the matrix elements $\mathcal{M}_{\Lambda^\ast}$ contribute to $\Lambda_b\to\Lambda^\ast(\to N\!\bar{K})\ell^+\ell^-$ decay through the following interference terms
\begin{equation}
\Gamma_2(s^a_\Lst,s^b_\Lst) = \frac{\sqrt{r_+ r_-}}{16 m_\Lst \pi^3} \sum_{s_N} \mathcal{M}_{\Lst}(s^a_\Lst,s_N) [\mathcal{M}_{\Lst}(s^b_\Lst,s_N)]^\ast\, ,
\end{equation}
where 
\begin{equation}
r_\pm = (m_\Lst \pm m_N)^2 - m_{\bar{K}}^2\, .
\end{equation}
Using the solutions of Rarita-Schwinger and Dirac fields given in appendix~\ref{app:Rarita} we obtain 
\begin{align}
&\Gamma_2(s^a_\Lst,s^b_\Lst) = \frac{\Gamma(\Lambda^\ast\to  N\!\bar{K})}{4} \times\, \nn\\& 
\small{\begin{pmatrix}
6 \sin ^2(\theta_\Lst ) & 2 \sqrt{3} e^{-i \phi } \sin (2 \theta_\Lst ) &
-2 \sqrt{3} e^{-2 i \phi } \sin ^2(\theta_\Lst ) & 0 \\
2 \sqrt{3} e^{i \phi } \sin (2 \theta_\Lst ) & 3 \cos (2 \theta_\Lst )+5 &
0 & -2 \sqrt{3} e^{-2 i \phi } \sin ^2(\theta_\Lst ) \\
-2 \sqrt{3} e^{2 i \phi } \sin ^2(\theta_\Lst ) & 0 & 3 \cos (2 \theta_\Lst
)+5 & -2 \sqrt{3} e^{-i \phi } \sin (2 \theta_\Lst ) \\
0 & -2 \sqrt{3} e^{2 i \phi } \sin ^2(\theta_\Lst ) & -2 \sqrt{3} e^{i \phi }
\sin (2 \theta_\Lst ) & 6 \sin ^2(\theta_\Lst )
\end{pmatrix}\, , }
\end{align}
where the $\Lambda^\ast\to N\!\bar{K}$ decay width is defined as
\begin{equation}
\Gamma(\Lambda^\ast\to N\!\bar{K}) = \frac{1}{4}\sum_{s_\Lst} \Gamma_2(s_\Lst,s_\Lst)\, .
\end{equation}
\section{Angular distributions \label{sec:angular}}
The results of the previous sections yield the following four-fold angular distribution for $\Lb\to \Lst(\to N\!\bar{K})\ell^+\ell^-$ decay
\begin{align}\label{eq:fourfold}
\frac{d^4\mathcal{B}}{dq^2 d\cl d\cLst d\phi} &= \frac{3}{8\pi} \bigg[\bigg(K_{1c}\cl + K_{1cc}\ccl + K_{1ss}\ssl \bigg)\ccLst\, \nn\\ 
&~~~~+\bigg(K_{2c}\cl + K_{2cc}\ccl + K_{2ss}\ssl  \bigg)\ssLst\, \nn\\
&~~~~+\bigg(K_{3ss}\ssl \bigg)\ssLst\cos\phi + \bigg(K_{4ss}\ssl\bigg)\ssLst\sin\phi\cos\phi \,\nn\\
&~~~~+\bigg(K_{5s}\sl + K_{5sc}\sl\cl \bigg)\sLst\cLst\cos\phi\, \nn\\
&~~~~+\bigg(K_{6s}\sl + K_{6sc}\sl\cl \bigg)\sLst\cLst\sin\phi
\bigg] \, .
\end{align} 
The $K_{\{\cdots \}}$, where ${\{\cdots \}}=1c, \cdots 6sc$, are called the angular coefficients that can be written in terms of the transversity amplitudes. As the masses of the final states has been kept, we show the mass corrections of the order $\mathcal{O}(m_\ell/\sqrt{q^2})$ and $\mathcal{O}(m_\ell^2/q^2)$ and write the angular coefficients as
\begin{equation}
K_{\{\cdots \}} = \mathcal{K}_{\{\cdots \}} + \frac{m_\ell}{\sqrt{q^2}} \mathcal{K}^\prime_{\{\cdots \}} + \frac{m_\ell^2}{q^2} \mathcal{K}^{\prime\prime}_{\{\cdots \}}\, .
\end{equation}  
The detailed expressions of $\mathcal{K}_{\{\cdots\}}$, $\mathcal{K}_{\{\cdots\}}^\prime$ and $\mathcal{K}^{\prime\prime}_{\{\cdots\}}$ in terms of the transversity amplitudes are given in the appendix \ref{app:angular}. The terms $\mathcal{K}^\prime$ and $\mathcal{K}^{\prime\prime}$ are important if the final state leptons are heavy, for example $\tau$ leptons.
 
\section{Observables \label{sec:obs}}
By weighted angular integrals of the differential distribution \eqref{eq:fourfold} we obtain different observables as a liner combinations of the angular coefficients and as a function of $q^2$
\begin{equation}
\mathcal{O}[\omega] = \displaystyle\int \frac{d^4\mathcal{B}}{dq^2 d\cl d\cLst d\phi} \omega(q^2,\theta_\ell,\theta_\Lst,\phi) d\cl d\cLst d\phi\, .
\end{equation} 
We limit ourselves to the following `simple' observables 
\begin{itemize}
\item $\omega=1$ yields the simplest observable, differential branching ratio as a function of $q^2$
\begin{equation}
\frac{d\mathcal{B}}{dq^2} = \frac{1}{3}\bigg[K_{1cc} + 2K_{1ss} + 2K_{2cc} + 4K_{2ss} + 2K_{3ss}  \bigg]\, .
\end{equation}
\item The fraction of longitudinal polarization of the dilepton system 
\begin{equation}
F_L = 1 - \frac{2(K_{1cc} + 2K_{2cc} )}{K_{1cc} + 2(K_{1ss} + K_{2cc} + 2K_{2ss} + K_{3ss} ) }\, ,
\end{equation}
is obtained for the choice
\begin{equation}
\omega = \frac{2-5\ccl}{d\mathcal{B}/dq^2}\, \nn.
\end{equation}
\item The choice
\begin{equation}
\omega = \frac{\rm sgn[\cl]}{d\mathcal{B}/dq^2}\, ,
\end{equation}
yields the well known lepton forward-backward asymmetry
\begin{equation}
A^\ell_{\rm FB} = \frac{3(K_{1c} + 2K_{2c} )}{2\big[ K_{1cc} + 2(K_{1ss} + K_{2cc} + 2K_{2ss} + K_{3ss} ) \big] }\, .
\end{equation}

\end{itemize}
Following the decay $\Lambda_b\to\Lambda(\to N\pi)\ell^+\ell^-$ \cite{Das:2018iap}, two more observables, the hadronic side asymmetry $A^{\Lst}_{\rm FB}$, and the mixed asymmetry $A^{\ell\Lst}_{\rm FB}$ can be defined corresponding to the weight factors $\omega = {\rm sgn}(\cos\theta_\Lst)/d\mathcal{B}/dq^2$ and $\omega = {\rm sgn}(\cos\theta_\Lst \cos\theta_\ell)/d\mathcal{B}/dq^2$, respectively. Since the $\Lst$ decays through strong interaction, these observables vanish.

\section{$\Lb\to\Lst$ in HQET \label{sec:fffree}} 
The description of $\Lb\to\Lst$ transition involves fourteen $q^2$-dependent form factors. In the limit of low hadronic recoil, the number of independent form factors reduce as a consequence of the HQET spin symmetry \cite{Isgur:1989ed, Isgur:1990pm, Isgur:1989vq, Mannel:1990vg, Grinstein:2004vb}. In this section we derive the improved Isgur-Wise relations between the form factors to the leading order in $1/m_b$ and including $\mathcal{O}(\alpha_s)$ corrections.
 
\subsection{Improved Isgur-Wise relations}
The starting point is to use the heavy baryon velocity $v^\mu = p^\mu/m_\Lb$ to project the $b$-quark filed  on to its large spinor components $h_v = \slashed{v}h_v$ in terms of leading Isgur-Wise form factors $\xi_{1,2}$
\begin{eqnarray}\label{eq:HQETleading}
\langle\Lst(k,s_\Lst)|\bar{s} \Gamma b |\Lb(p=v m_\Lb, s_\Lb)\rangle &\to& \langle\Lst(k,s_\Lst)|\bar{s} \Gamma h_v |\Lb(v, s_\Lb)\rangle\,\nn\\ &\simeq& \bar{U}_\Lst^\alpha(k,s_\Lst) v_\alpha (\xi_1 + \slashed{v}\xi_2 ) \Gamma u_\Lb(v,s_\Lb)\, ,
\end{eqnarray}
where $U_\Lst^\alpha$ is the Rarita-Schwinger spinor describing the $\Lst$, $u_\Lb$ is a Dirac spinor describing the $\Lb$, and $\Gamma$ is a Dirac matrix. The Isgur-Wise form factors are defined as
\begin{equation}
\xi_{1,2} \equiv \xi_{1,2}(v.k)\, .
\end{equation}
In the limit of heavy quark $m_b\to \infty$, \emph{i.e.,} neglecting the contributions of the order $1/m_b$ in the parametrizations of the hadronic matrix elements \eqref{eq:ffVA1}-\eqref{eq:ffT} and comparing with \eqref{eq:HQETleading} we get 
\begin{align}
& f^V_\perp = f^V_0 = f^A_t = f^T_\perp = f^T_0 = \frac{\xi_1 - \xi_2}{m_\Lb}\, ,\\
& f^A_\perp = f^A_0 = f^V_t = f^{T5}_\perp = f^{T5}_0 = \frac{\xi_1 + \xi_2}{m_\Lb}\, ,\\
& f^V_g = f^A_g = f^T_g = f^{T5}_g = 0\, .
\end{align}

To account for the sub-leading corrections of the order $\mathcal{O}(\alpha_s)$, following \cite{Grinstein:2004vb} we use the matching relations of the QCD currents onto the HQET. For the (axial-)vector currents the matching relations read
\begin{eqnarray}
\bar{s} \gamma^\mu b &=& C_0^{(v)} \bar{s}\gamma^\mu h_v + C_1^{(v)} v^\mu \bar{s} h_v + \frac{1}{2m_b} \bar{s} \gamma^\mu i \slashed{D}_\perp h_v + \cdots\,,\\
\bar{s} \gamma^\mu \gamma_5 b &=& C_0^{(v)} \bar{s}\gamma^\mu\gamma_5 h_v - C_1^{(v)} v^\mu \bar{s} \gamma_5 h_v - \frac{1}{2m_b} \bar{s} \gamma^\mu i \slashed{D}_\perp \gamma_5 h_v + \cdots\,,
\end{eqnarray}
and for the (pseudo-)tensor currents the relation is
\begin{equation}
\bar{s} i \sigma^{\mu\nu} q_\nu(\gamma_5) b = C_0^{(t)} \bar{s} i \sigma^{\mu\nu} q_\nu(\gamma_5)h_v \pm \frac{1}{2m_b} \bar{s}\sigma^{\mu\nu} q_\nu i\slashed{D}_\perp (\gamma_5) h_v + \cdots\, ,
\end{equation} 
where 
\begin{equation}
D_\mu = v_\mu(v.D)+D_{\perp\mu}\, ,\quad D_{\perp\mu} = (g_{\mu\nu}-v_\mu v_\nu)D^\nu\, .\nn
\end{equation}
The renormalization scale $\mu$-dependent matching coefficients $C_0^{(v,t)}, C_1^{(v)}$ at next-to-leading order in $\alpha_s$ are \cite{Grinstein:2004vb}
\begin{align}
&C_0^{(v)} = 1 - \frac{\alpha_s C_F}{4\pi} \bigg( 3\ln\bigg( \frac{\mu}{m_b} \bigg) + 4 \bigg) + \mathcal{O}(\alpha_s^2)\, ,\\
&C_1^{(v)} = \frac{\alpha_s C_F}{2\pi} + \mathcal{O}(\alpha_s^2)\, ,\\
& C_0^{(t)} = 1 - \frac{\alpha_s C_F}{4\pi} \bigg( 5\ln\bigg( \frac{\mu}{m_b} \bigg) + 4 \bigg) + \mathcal{O}(\alpha_s^2)\, .
\end{align}
The matrix elements of these currents can be parametrized in terms of the leading Isgur-Wise form factors $\xi_{1,2}$ as
\begin{align}\label{eq:ffparam2}
&\langle \Lst(k,s_\Lst) | \bar{s} \gamma^\mu(\gamma_5) b |\Lb(p=m_\Lb v,s_\Lb)\nn \\&\quad\quad\quad \simeq C_0^{(v)} \sum_{n=1,2} \xi_n U_\Lst^{\alpha} (k,s_\Lst) v_\alpha \Gamma_n \gamma^\mu (\gamma_5) u_\Lb(v,s_\Lb) \, \nn\\
&\quad\quad\quad \pm C_1^{(v)} \sum_{n=1,2} \xi_n v^\mu U_\Lst^{\alpha} (k,s_\Lst) v_\alpha \Gamma_n (\gamma_5) u_\Lb(v,s_\Lb)\, ,~~~~~~\\
\label{eq:ffparam3}
&\langle \Lst(k,s_\Lst) | \bar{s} i \sigma^{\mu\nu} q_\nu (\gamma_5) b |\Lb(p=m_\Lb v,s_\Lb)\nn \\ &\quad\quad\quad \simeq C_0^{(t)} \sum_{n=1,2} \xi_n U_\Lst^{\alpha} (k,s_\Lst) v_\alpha \Gamma_n i \sigma^{\mu\nu}q_\nu (\gamma_5) u_\Lb(v,s_\Lb) \, ,
\end{align}
where the two independent Dirac structures are 
\begin{equation}
\Gamma_1 = 1\, ,\quad \Gamma_2  = \slashed{v}\, .
\end{equation}
Comparing the parametrizations \eqref{eq:ffparam2} and \eqref{eq:ffparam3} with \eqref{eq:ffVA1}-\eqref{eq:ffT} we get the following expressions for the physical form factors at leading order at $1/m_b$ and including $\mathcal{O}(\alpha_s)$ corrections 
\begin{align}\label{eq:IW1}
&f_\perp^{V,A} = C_0^{(v)} \frac{(\xi_1\mp\xi_2)}{m_\Lb}\, ,\\
&f_0^{V,A} = \bigg( C_0^{(v)} + \frac{C_1^{(v)}s_\pm}{2m_\Lb(m_\Lb\pm m_\Lst)} \bigg)\frac{\xi_1}{m_\Lb} \mp \bigg( C_0^{(v)} - \frac{(2C_0^{(v)} + C_1^{(v)} )s_\pm}{2m_\Lb(m_\Lb\pm m_\Lst)} \bigg) \frac{\xi_2}{m_\Lb} \, ,\\
& f_\perp^{T(5)} = C_0^{(t)}\bigg( \frac{(\xi_1\mp\xi_2)}{m_\Lb}  \pm \frac{s_\pm}{m_\Lb(m_\Lb\pm m_\Lst)} \frac{\xi_2}{m_\Lb} \bigg)\, ,\\
& f_0^{T(5)} = C_0^{(t)} \frac{(\xi_1\mp\xi_2)}{m_\Lb}\, ,\\
&f_t^V(q^2) = \frac{1}{m_{\Lambda_b}}\xi_1\bigg(C_0^{(v)}+C_1^{(v)}\Big(1-\frac{s_-}{2m_{\Lambda_b}(m_{\Lambda_b}-m_{\Lambda^{\star}})}\Big)\bigg)\nn\\
&\quad\quad\quad+\frac{1}{m_{\Lambda_b}}\xi_2\bigg(C_0^{(v)}\Big(1-\frac{s_-}{m_{\Lambda_b}(m_{\Lambda_b}-m_{\Lambda^{\star}})}\Big)+C_1^{(v)}\Big(1-\frac{s_-}{2m_{\Lambda_b}(m_{\Lambda_b}-m_{\Lambda^{\star}})}\Big)\bigg), \\
\label{eq:IW2}
&f_t^A(q^2) = \frac{1}{m_{\Lambda_b}}\xi_1\bigg(C_0^{(v)}+C_1^{(v)}\Big(1-\frac{s_+}{2m_{\Lambda_b}(m_{\Lambda_b}+m_{\Lambda^{\star}})}\Big)\bigg)\nn\\
&\quad\quad\quad-\frac{1}{m_{\Lambda_b}}\xi_2\bigg(C_0^{(v)}\Big(1-\frac{s_+}{m_{\Lambda_b}(m_{\Lambda_b}+m_{\Lambda^{\star}})}\Big)+C_1^{(v)}\Big(1-\frac{s_+}{2m_{\Lambda_b}(m_{\Lambda_b}+m_{\Lambda^{\star}})}\Big)\bigg)\, .\
\end{align}
The form factors $f^{V,A}_g$ remain zero. These expressions will be used to correlate the form factors and reduce the number of independent form factors in the transversity amplitudes. 

\subsection{Low-recoil factorization \label{sec:lowfac}}
The improved Isgur-Wise relations \eqref{eq:IW1}-\eqref{eq:IW2} lead to simplifications of the description of the decay at low recoil region. In what follows, we consider the (axial-)vector form factors $f^{V,A}_0$, $f^{V,A}_\perp$, $f^{V,A}_t$ as independent and use the improved Isgur-Wise relations \eqref{eq:IW1}-\eqref{eq:IW2} to relate the tensor and pseudo-tensor form factors. With this consideration, at leading order in $1/m_b$ and including $\mathcal{O}(\alpha_s)$ corrections, the (axial-)vector type transversity amplitudes $B_{\perp1}^{L(R)}$, $B_{\|1}^{L(R)}$ vanish, and $A_{\perp1,0}^{L(R)}$ and $A_{\|1,0}^{L(R)}$ depend on one form factors each
\begin{align}\label{eq:TAsHQET}
& B_{\perp 1}^{L(R)} = 0\, ,\quad B_{\| 1}^{L(R)} = 0\, ,\\
& A^{L(R)}_{\perp 0} = -\sqrt{2}N \frac{m_\Lb + m_\Lst}{\sqrt{q^2}} \frac{s_- \sqrt{s_+}}{\sqrt{6}m_\Lst} \mC_+^{L(R)} f_0^V\, ,\\
&A^{L(R)}_{\|0} =  \sqrt{2}N \frac{m_\Lb - m_\Lst}{\sqrt{q^2}} \frac{s_+ \sqrt{s_-}}{\sqrt{6}m_\Lst} \mC_-^{L(R)} f_0^A\, ,\\
&A^{L(R)}_{\perp 1} = -\sqrt{2}N \frac{s_- \sqrt{s_+}}{\sqrt{3}m_\Lst} \mC_+^{L(R)} f_\perp^V\, ,\\
\label{eq:TAsHQETend}
&A^{L(R)}_{\|1} =  -\sqrt{2}N \frac{s_+ \sqrt{s_-}}{\sqrt{3}m_\Lst} \mC_-^{L(R)} f_\perp^A\, ,
\end{align}
where the combinations of Wilson coefficients are 
\begin{align}
\begin{split}\label{eq:Cpm}
& \mC_+^{L(R)} = (\mC_9 + \mC_{9^\prime}) + \frac{2\kappa m_b m_\Lb}{q^2} (\mC_7 + \mC_{7^\prime}) \mp (\mC_{10} + \mC_{10^\prime})\, ,\quad\\
& \mC_-^{L(R)} = (\mC_9 - \mC_{9^\prime}) + \frac{2\kappa m_b m_\Lb}{q^2} (\mC_7 - \mC_{7^\prime}) \mp (\mC_{10} - \mC_{10^\prime})\, .
\end{split}
\end{align}
The parameter 
\begin{equation}
\kappa \equiv \kappa(\mu) = \frac{C_0^{(t)}}{C_0^{(v)}} =1-\frac{\alpha_sC_F}{2\pi} \ln\bigg( \frac{\mu}{m_b}  \bigg)\, ,
\end{equation}
absorbs the perturbative QCD corrections to the form factor relations in such a way that with the product of the Wilson coefficients and the $b$-quark mass, the transversity amplitudes are free of renormalization-scale at a given order in the perturbation theory. In this derivation, we have ignored the sub-leading terms of the order $m_\Lst/m_\Lb$ and $\Lambda_{\rm QCD}/m_\Lb$, and used a naively anti-commutating $\gamma_5$ matrix. The consequences of the simplified expressions of the tranversity amplitudes translate to the factorization of short- and long-distance physics in the decay observables. The factorization in scale makes the electroweak physics transparent in the observables. In the subsequent subsections, the low-recoil factorization is discussed. For these discussions, we will neglect the mass of the leptons which is a valid approximation in the low-recoil region if the leptons are muons.

\subsubsection{In SM basis}
In the SM+SM$^\prime$ set of operators the independent short-distance coefficients are
\begin{align} \label{eq:SD}
\rho_1^\pm =\frac{1}{2} \left(|C_\pm^R|^2 +|C_\pm^L|^2\right) \, ,  \quad 
\rho_2^\pm  =\frac{1}{4} \left(C_+^R C_-^{R*} \mp C_-^L C_+^{L*}\right) \, .
\end{align}
The coefficients $\rho_1^\pm$ and $\rho_2^+$ also appear in $B\to K^\ast\ell^+\ell^-$ \cite{Bobeth:2010wg} and $B\to K\pi\ell^+\ell^-$ \cite{Das:2014sra} decay, and $\rho_2^-$ appear in $\Lb\to\Lambda(\to N\pi)\ell^+\ell^-$ \cite{Boer:2014kda} decay.
In the SM ($\mC_{9^\prime, 10^\prime}=0$  and $\mC_{S^{(\prime)}, P^{(\prime)}}=0$)  
\begin{equation}
\mC^{L(R)} \equiv \mC^{L(R)}_+ = \mC^{L(R)}_- =  \mC_9 + \frac{2\kappa m_b m_\Lb}{q^2} \mC_7 \mp \mC_{10}\, ,
\end{equation}
and therefore, in the SM observables only two short-distance coefficients $\rho_{1,2}$ are relevant \cite{Das:2014sra}
\begin{align} \label{eq:SMrho}
\rho_1 \equiv \rho_1^\pm=2 {\rm Re} \rho_2^- \, , \quad \rho_2 \equiv {\rm Re} \rho_2^+ \, , \quad
{\rm Im} \rho_2^\pm=0\, . \hspace{1cm} 
\end{align}
The low recoil factorization of the angular coefficients yields
\begin{eqnarray}
K_{1c} &=& \frac{8N^2}{3m_\Lst^2}(s_+s_-)^{3/2}f_\perp^Vf_\perp^A \rho_2\, ,\\
K_{1cc}&=&\frac{2N^2}{3m_\Lst^2}s_+s_-\bigg(s_-|f_\perp^V|^2 + s_+|f_\perp^A|^2 \bigg) \rho_1\, ,\\
K_{1ss}&=&\frac{N^2}{3m_\Lst^2}s_+s_-\bigg(\frac{(m_\Lb+m_\Lst)^2}{q^2}s_- |f_0^V|^2 + \frac{(m_\Lb-m_\Lst)^2}{q^2}s_+ |f_0^A|^2\nn\\
&+&s_- |f_\perp^V|^2 + s_+ |f_\perp^A|^2 \bigg) \rho_1\, ,\\
K_{2c}&=&\frac{2N^2}{3m_\Lst^2}(s_+s_-)^{3/2}f_\perp^Vf_\perp^A \rho_2\, ,\\
K_{2cc}&=&\frac{N^2}{6m_\Lst^2}s_+s_-\bigg(s_- |f_\perp^V|^2 + s_+ |f_\perp^A|^2 \bigg) \rho_1\, ,\\
K_{2ss}&=&\frac{N^2}{12m_\Lst^2}s_+s_-\bigg(\frac{(m_\Lb+m_\Lst)^2}{q^2}s_- |f_0^V|^2 + \frac{(m_\Lb-m_\Lst)^2}{q^2}s_+ |f_0^A|^2\nn\\
&+&s_-|f_\perp^V|^2+ s_+ |f_\perp^A|^2 \bigg) \rho_1 \, .
\end{eqnarray}
The rest of the coefficients vanish due to \eqref{eq:TAsHQET}. 

The simple observables $d\mathcal{B}/dq^2$, $A_{\rm FB}^\ell$ and $F_L$ factorize into long- and short-distance physics as
\begin{align}
&\frac{d\mathcal{B}}{dq^2} = \frac{N^2s_+s_-}{3m_\Lst^2}\bigg[s_- \Big(2 |f_\perp^V|^2 + \frac{(m_\Lb+m_\Lst)^2}{q^2} |f_0^V|^2 \Big)\, \nn\\ &\quad\quad+ s_+ \Big(2|f_\perp^A|^2 + \frac{(m_\Lb+m_\Lst)^2}{q^2} |f_0^A|^2 \Big)\bigg] \rho_1 \, ,\\
&\frac{d\mathcal{B}}{dq^2}F_L= \frac{d\mathcal{B}}{dq^2} - \frac{2N^2}{3\mmL}s_+s_-\Bigg(s_- |f_\perp^V|^2+ s_+ |f_\perp^A|^2\Bigg) \rho_1 \, ,\\
&\frac{d\mathcal{B}}{dq^2}A_{\rm FB}^\ell=\frac{2N^2}{\mmL}(s_+s_-)^{3/2} f_\perp^Vf_\perp^A \rho_2 \, .
\end{align}

The merit of low-recoil factorization is that it lets us construct important tests in the SM. For example, the following ratios of angular coefficients are constants
\begin{align}
&\frac{K_{1c}}{K_{2c}} = 4\, ,\quad \frac{K_{1cc}}{K_{2cc}}=4\, ,\quad \frac{K_{1ss}}{K_{2ss}} = 4\, . 
\end{align}
The following ratios depend on ratios of form factor and a short-distance ratio $\rho_2/\rho_1$ 
\begin{equation}
\frac{K_{1c}}{K_{1cc}} = \frac{K_{2c}}{K_{2cc}} =   \Bigg( \frac{4\sqrt{s_+s_-}}{ s_- \frac{f^V_\perp}{f^A_\perp} + s_+ \frac{f^A_\perp}{f^V_\perp} }  \Bigg) \frac{\rho_2}{\rho_1}\, .
\end{equation}
If the form factors are known, then the ratios can be used to extract $\rho_2/\rho_1$. We also construct the following ratios that are independent of short-distance physics in the SM
\begin{equation}
	\frac{K_{1ss}}{K_{1cc}} = \frac{K_{2ss}}{K_{2cc}} = \frac{1}{2}\bigg[1+\frac{(m_\Lb+m_\Lst)^2 s_-|f_0^V|^2 + (m_\Lb-m_\Lst)^2 s_+|f_0^A|^2}{q^2 \big(s_-|f_\perp^V|^2+s_+|f_\perp^A|^2 \big)}\bigg]\, .
\end{equation}
These can be used to test form factors using experimental data.
\subsubsection{In SM+SM$^\prime$+SP}
Going beyond the SM operator basis, when we include SM$^\prime$+SP set of operators all the four short-distance coefficients $\rho_{1,2}^\pm$ contribute. The low-recoil factorization yields 
\begin{align}
& K_{1c} =  \frac{8N^2}{3m_\Lst^2}(s_+s_-)^{3/2}f_\perp^Vf_\perp^A \re\rho_2^+\, ,\\
& K_{1cc} = \frac{2N^2}{3m_\Lst^2}s_+s_-\bigg(s_-\rho_1^+|f_\perp^V|^2+s_+\rho_1^-|f_\perp^A|^2\nn\\\
&\quad\quad+\frac{1}{2}\Big[\frac{(m_\Lb-m_\Lst)^2}{(m_b-m_s)^2}s_+\rho_{\rm S}^+|f_t^V|^2+ \frac{(m_\Lb+m_\Lst)^2}{(m_b+m_s)^2}s_-\rho_{\rm S}^-|f_t^A|^2\Big]\bigg)\, ,\\
&K_{1ss}=\frac{N^2}{3m_\Lst^2}s_+s_-\bigg(\frac{(m_\Lb+m_\Lst)^2}{q^2}s_-\rho_1^+|f_0^V|^2+\frac{(m_\Lb-m_\Lst)^2}{q^2}s_+\rho_1^-|f_0^A|^2\nn\\
&\quad\quad +s_-\rho_1^+|f_\perp^V|^2+s_+\rho_1^-|f_\perp^A|^2+\frac{(m_\Lb-m_\Lst)^2}{(m_b-m_s)^2}s_+\rho_{\rm S}^+|f_t^V|^2\, \nn\\&\quad\quad+ \frac{(m_\Lb+m_\Lst)^2}{(m_b+m_s)^2}s_-\rho_{\rm S}^-|f_t^A|^2\bigg)\, ,\\
&K_{2c}=\frac{2N^2}{3m_\Lst^2}(s_+s_-)^{3/2}f_\perp^Vf_\perp^A \re\rho_2^+\, ,\\
&K_{2cc}=\frac{N^2}{6m_\Lst^2}s_+s_-\bigg(s_-\rho_1^+|f_\perp^V|^2+s_+\rho_1^-|f_\perp^A|^2\nn\\\
&\quad\quad +\frac{1}{2} \Big[\frac{(m_\Lb-m_\Lst)^2}{(m_b-m_s)^2}s_+\rho_{\rm S}^+|f_t^V|^2+ \frac{(m_\Lb+m_\Lst)^2}{(m_b+m_s)^2}s_-\rho_{\rm S}^-|f_t^A|^2\Big]\bigg)\, ,\\
&K_{2ss}=\frac{N^2}{12m_\Lst^2}s_+s_-\bigg(\frac{(m_\Lb+m_\Lst)^2}{q^2}s_-\rho_1^+|f_0^V|^2+\frac{(m_\Lb-m_\Lst)^2}{q^2}s_+\rho_1^-|f_0^A|^2\nn\\ &\quad\quad+s_-\rho_1^+|f_\perp^V|^2+s_+\rho_1^-|f_\perp^A|^2+\frac{(m_\Lb-m_\Lst)^2}{(m_b-m_s)^2}s_+\rho_{S}^+|f_t^V|^2\, \nn\\&\quad\quad+ \frac{(m_\Lb+m_\Lst)^2}{(m_b+m_s)^2}s_-\rho_{S}^-|f_t^A|^2\bigg)\, .
\end{align}
Here we have defined new short-distance coefficients that encode the scalar NP
\begin{equation}
\rho_{\rm S}^{\pm} =\frac{1}{2} \left(|\mC_{\rm SP\pm}^R|^2 +|\mC_{\rm SP\pm}^L|^2\right)\, .
\end{equation}

In this extended set of operators, the simple observables $d\mathcal{B}/dq^2$, $A_{\rm FB}^\ell$ and $F_L$ read 
\begin{align}
&\frac{d\mathcal{B}}{dq^2} = \frac{N^2}{9m_\Lst^2}s_+s_-\Bigg(3\Big[\frac{(m_\Lb+m_\Lst)^2}{q^2}s_-\rho_1^+|f_0^V|^2+\frac{(m_\Lb-m_\Lst)^2}{q^2}s_+\rho_1^-|f_0^A|^2\Big]\nn\\\
&\quad\quad+6\Big[s_-\rho_1^+|f_\perp^V|^2+s_+\rho_1^-|f_\perp^A|^2\Big]+\frac{9}{2}\Big[\frac{(m_\Lb-m_\Lst)^2}{(m_b-m_s)^2}s_+\rho_{S}^+|f_t^V|^2\,\nn\\&\quad\quad+ \frac{(m_\Lb+m_\Lst)^2}{(m_b+m_s)^2}s_-\rho_{S}^-|f_t^A|^2\Big]\Bigg)\, ,\\
&\frac{d\mathcal{B}}{dq^2}F_L= \frac{d\mathcal{B}}{dq^2} - \frac{N^2s_+s_-}{9m_\Lst^2} \Bigg(6s_-\rho_1^+|f_\perp^V|^2+6s_+\rho_1^-|f_\perp^A|^2+ 3 \frac{(m_\Lb-m_\Lst)^2}{(m_b-m_s)^2}s_+\rho_{S}^+|f_t^V|^2\, \nn\\&\quad\quad\quad+3 \frac{(m_\Lb+m_\Lst)^2}{(m_b+m_s)^2}s_-\rho_{S}^-|f_t^A|^2 \Bigg)\, ,\\
&\frac{d\mathcal{B}}{dq^2}A_{\rm FB}^\ell=\frac{2N^2}{m_\Lst^2}(s_+s_-)^{3/2}f_\perp^Vf_\perp^A \re\rho_2^+\, .
\end{align}

For the ratios of angular coefficients constructed in the previous section we make the following observations 
\begin{itemize}
\item Interestingly, the ratios $K_{1c}/K_{2c}$, $K_{1cc}/K_{2cc}$ and $K_{1ss}/K_{2ss}$ remain independent of both short- and long-distance physics in the extended set of operators.
\item If only SM$^\prime$ NP is present, then both $K_{1c}/K_{1cc}$ and $K_{2c}/K_{2cc}$ are sensitive to it. Irrespective of the presence of SM$^\prime$ NP, the ratios are sensitive to scalar NP.
\item For $K_{1ss}/K_{1cc}$ and $K_{2ss}/K_{2cc}$ the dependence on the new physics follow the same pattern as in $K_{1c}/K_{1cc}$ and $K_{2c}/K_{2cc}$.
\end{itemize}

\section{Numerical Analysis  \label{sec:num}}
In this section, we perform a numerical analysis of the $\Lb\to\Lst(\to N\!\bar{K})\mu^+\mu^-$ observables and study their sensitivity to the SM and NP. Such an analysis require the knowledge of the form factors. At present, the lattice QCD calculations of the form factors are only preliminary \cite{Meinel:2016cxo}. Our numerical analysis is based on the non-relativistic quark model predictions (using the ``full quark model wave function (MCN)'' model) presented in \cite{Mott:2011cx}. We assume uncorrelated 30\% uncertainties on the four contributing form factors $f_0^{V,A}, f_\perp^{V,A}$ for illustrative purpose in the absence of such in Ref.~\cite{Mott:2011cx}. To account for the uncertainties due to the neglected terms of the order $\mathcal{O}(\Lambda/m_b)$ and of the order $\mathcal{O}(m_\Lst/m_\Lb)$ in the improved Isgur-Wise relations, we assume 10\% corrections to the amplitudes. These corrections are included by multiplying the amplitudes $A_{\perp,\|0}$, $A_{\perp,\|1}$ by uncorrelated real scale factors. 

Even if the form factors are precisely known, there are theoretical uncertainties due to purely hadronic operators $\mathcal{O}_{1\cdots 6}$, and penguin operators $\mathcal{O}_8$, combined with a virtual photon emissions. These are non-local effects because the electromagnetic vertex is separated from the quark flavor transition by a characteristic distance that is quite large. At low-recoil, a tool developed to calculate these long-distance contributions in terms of short-distance effects combine the HQET with low-recoil OPE in $1/Q$ with $Q\sim (m_b, \sqrt{q^2})$ \cite{Grinstein:2004vb}. It allows us to calculate the non-local matrix elements in terms of local matrix elements in the powers of $1/Q$. The contributions are absorbed in the Wilson coefficients $\mC_{7,9}$ and are renamed $\mC_{7,9}^{\rm eff}$, the expressions of which are given in appendix of \cite{Das:2018iap}. The SM values of all the Wilson coefficients are taken from \cite{Altmannshofer:2008dz}, evaluated at the $b$-quark mass scale $\mu=m_b = 4.8$ GeV. The $b$- and the $c$-quark loop functions are taken from \cite{Buras:1994dj, Misiak:1992bc, Beneke:2001at, Seidel:2004jh}. The terms that are neglected in the OPE are of the order $\mathcal{O}(\alpha_s\Lambda/m_b, m_c^4/Q^4)$. We assume 5\% corrections of the order $\mathcal{O}(\alpha_s \Lambda/m_b)$ to the amplitudes due to the neglected terms. The corrections are included by multiplying the amplitudes $A^{L(R)}_{\perp,\|0}$, $A^{L(R)}_{\perp,\| 1}$ by uncorrelated real scale factors. 
 
The most important non-local effects come from charm quarks that form charm-loop to which the virtual photon can be attached. These effects are common to all $b\to s\ell\ell$ transitions. Above the $c\bar{c}$ threshold, the charm-loops give rise to a resonance spectrum. In experiments, the narrow resonances $J/\Psi$ and $\psi(2S)$ are cut off from the spectrum and therefore may not be relevant in the low-recoil region. However, there still remain few broad charmonium resonances  that have contributions in this region and are responsible for the violation of the quark-hadron duality \cite{Lyon:2014hpa}. These resonances are not captured at any order in the OPE. In the $B\to K\ell^+\ell^-$ decay, the duality violation in the integrated decay rate over the high-$q^2$ region is estimated to be around 2\% \cite{Beylich:2011aq}. Such an analysis for the present mode is beyond the scope of this paper and we ignore it. Let us also emphasize that the low-recoil OPE does not capture the local resonances at any order in the expansion and hence, the actual $q^2$ distributions may be locally away from the OPE predictions.  

\begin{figure}[h!]
	\begin{center}
		\includegraphics[scale=0.29]{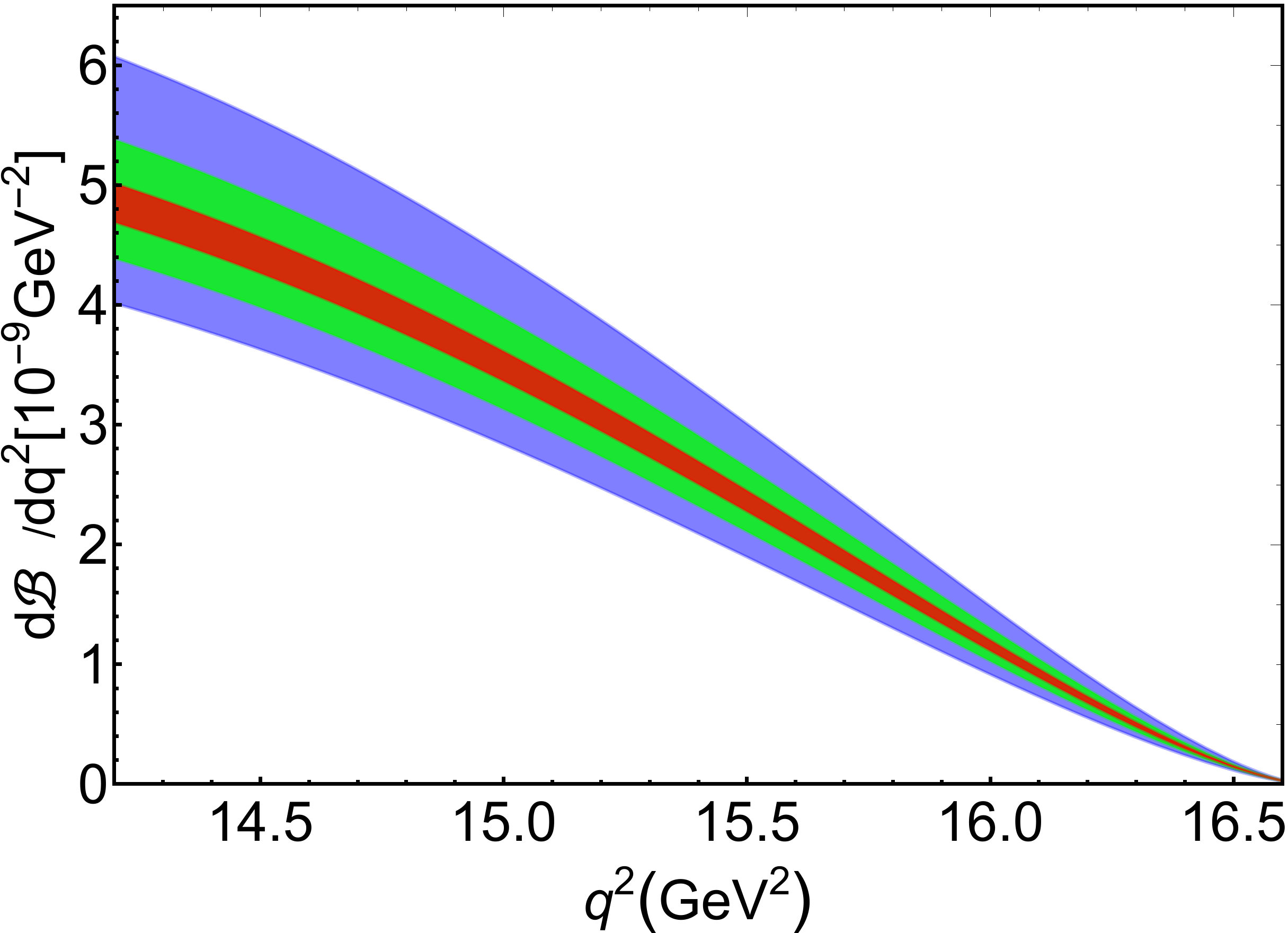}
		\includegraphics[scale=0.304]{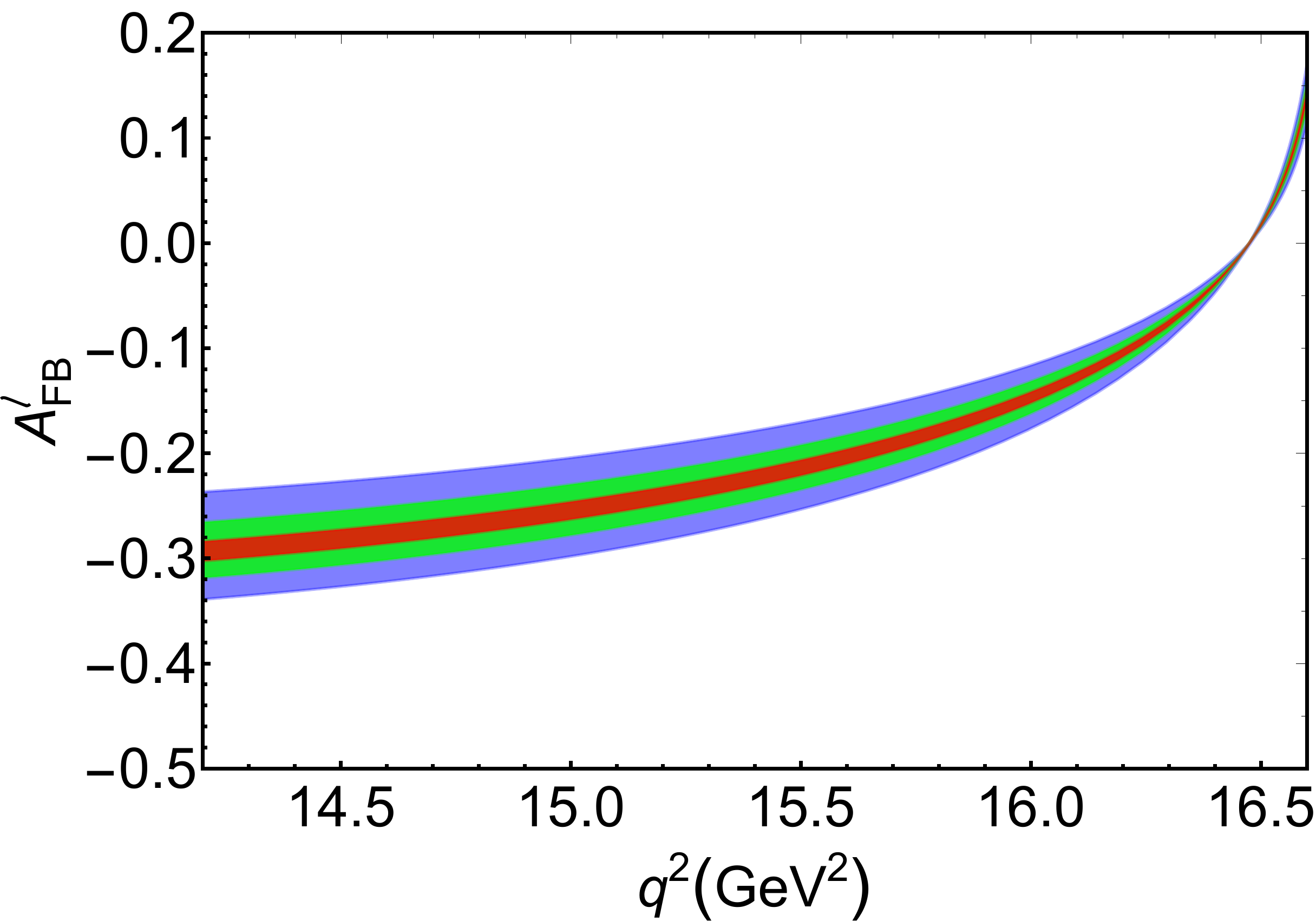}
		\includegraphics[scale=0.3]{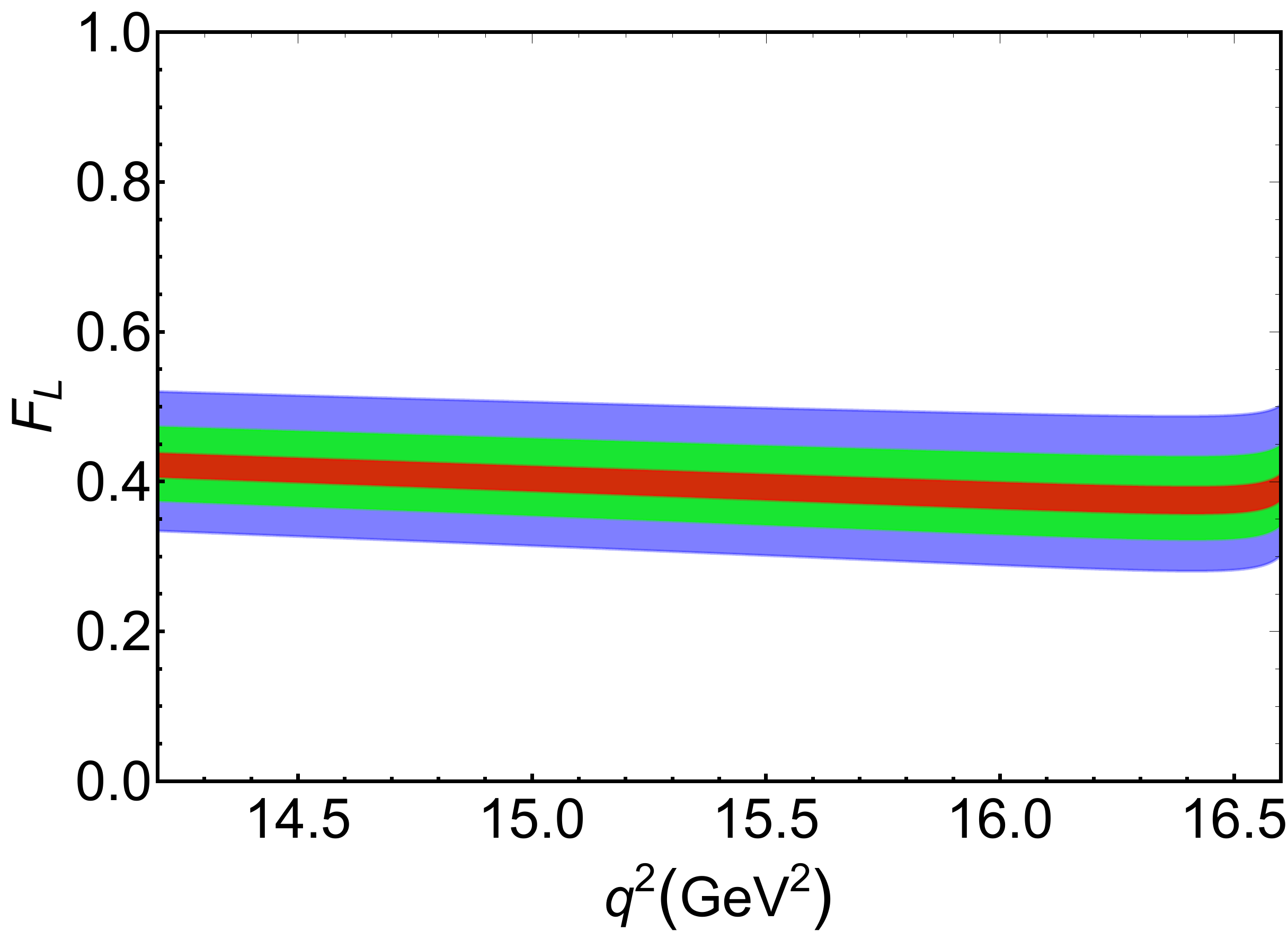}%
		\caption{The differential branching ratio, lepton-side forward-backward asymmetry, and the longitudinal polarization fraction of $\Lb\to\Lst(\to N\bar{K})\mu^+\mu^-$ in the SM at the low-recoil. The blue, green, and red bands correspond to the uncertainties coming from the form factors, corrections to the Isgur-Wise relations, and sub-leading corrections to the amplitudes. The figures are for illustrative purpose only (see text for details). \label{fig:SM1}}
	\end{center}
\end{figure}

\begin{figure}[h!]
	\begin{center}
		\includegraphics[scale=0.3]{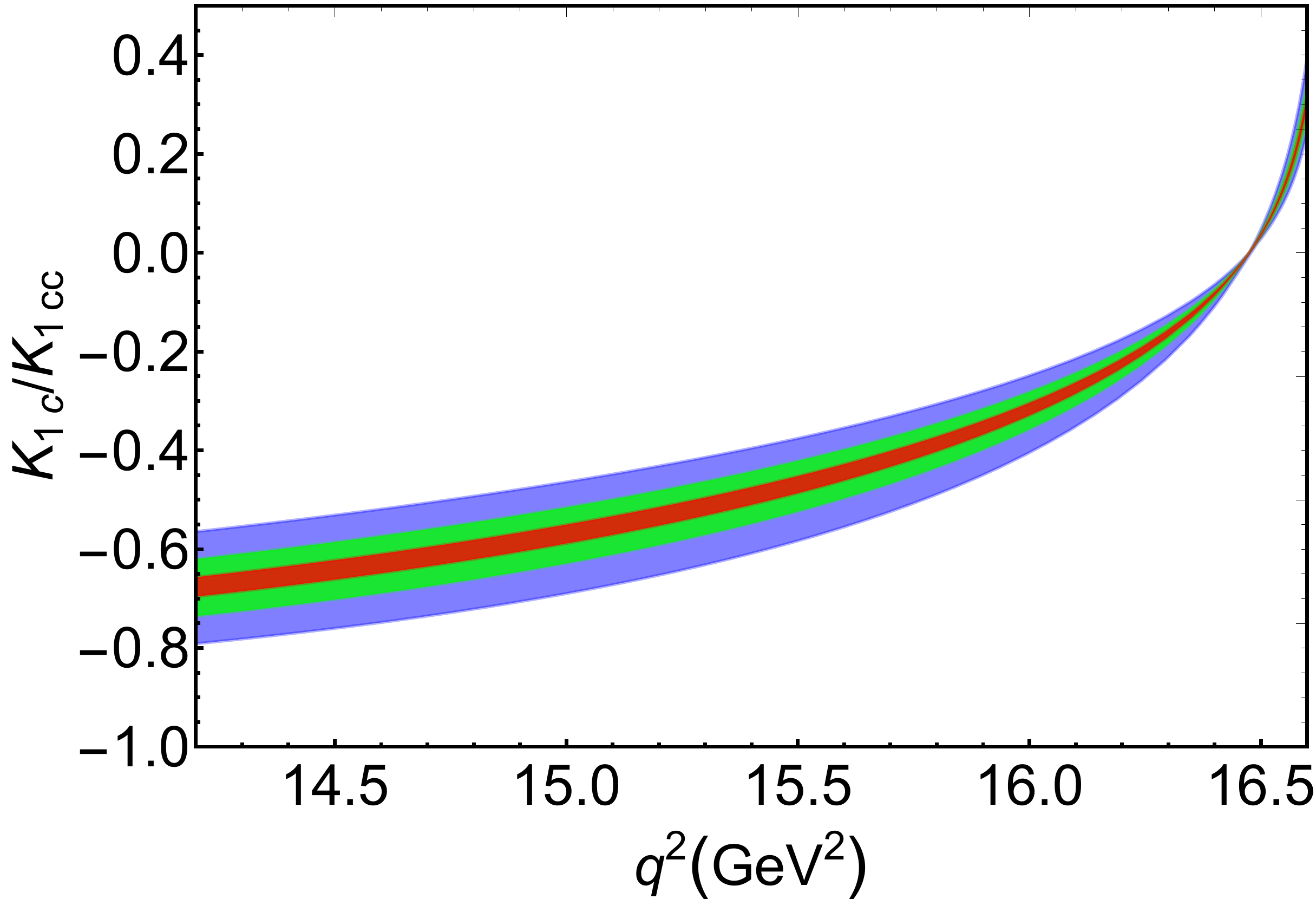}
		\includegraphics[scale=0.3]{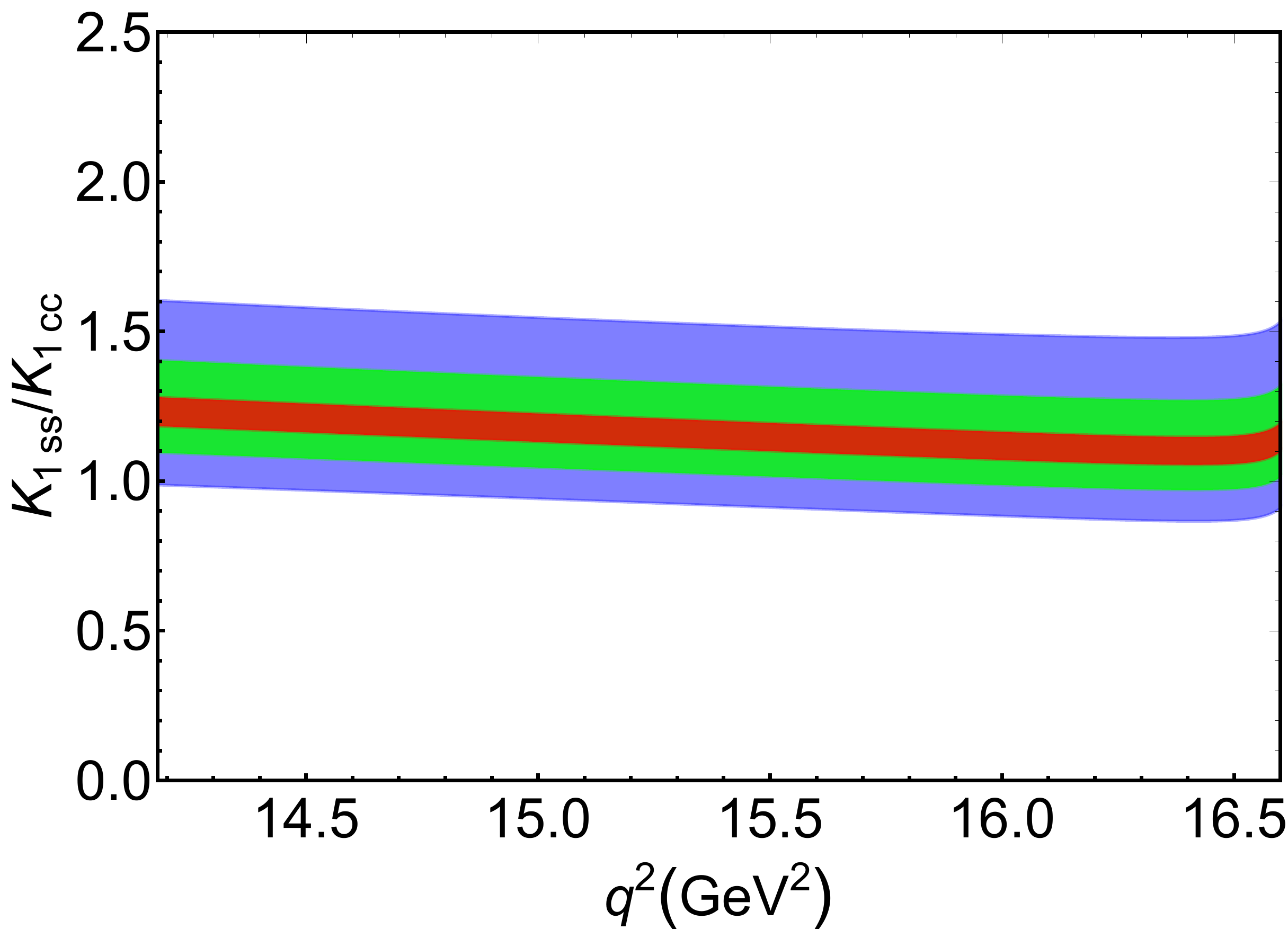}
		\caption{The SM predictions of $K_{1c}/K_{1cc}$ and $K_{1ss}/K_{1cc}$ at low-recoil for $\Lb\to\Lst(\to N\bar{K})\mu^+\mu^-$. The meaning of the bands are same as in the Fig.~\ref{fig:SM1}. The figures are for illustrative purpose only (see text for details). \label{fig:RatSMlow}}
	\end{center}
\end{figure}

\begin{figure}[h!]
	\begin{center}
		\includegraphics[scale=0.3]{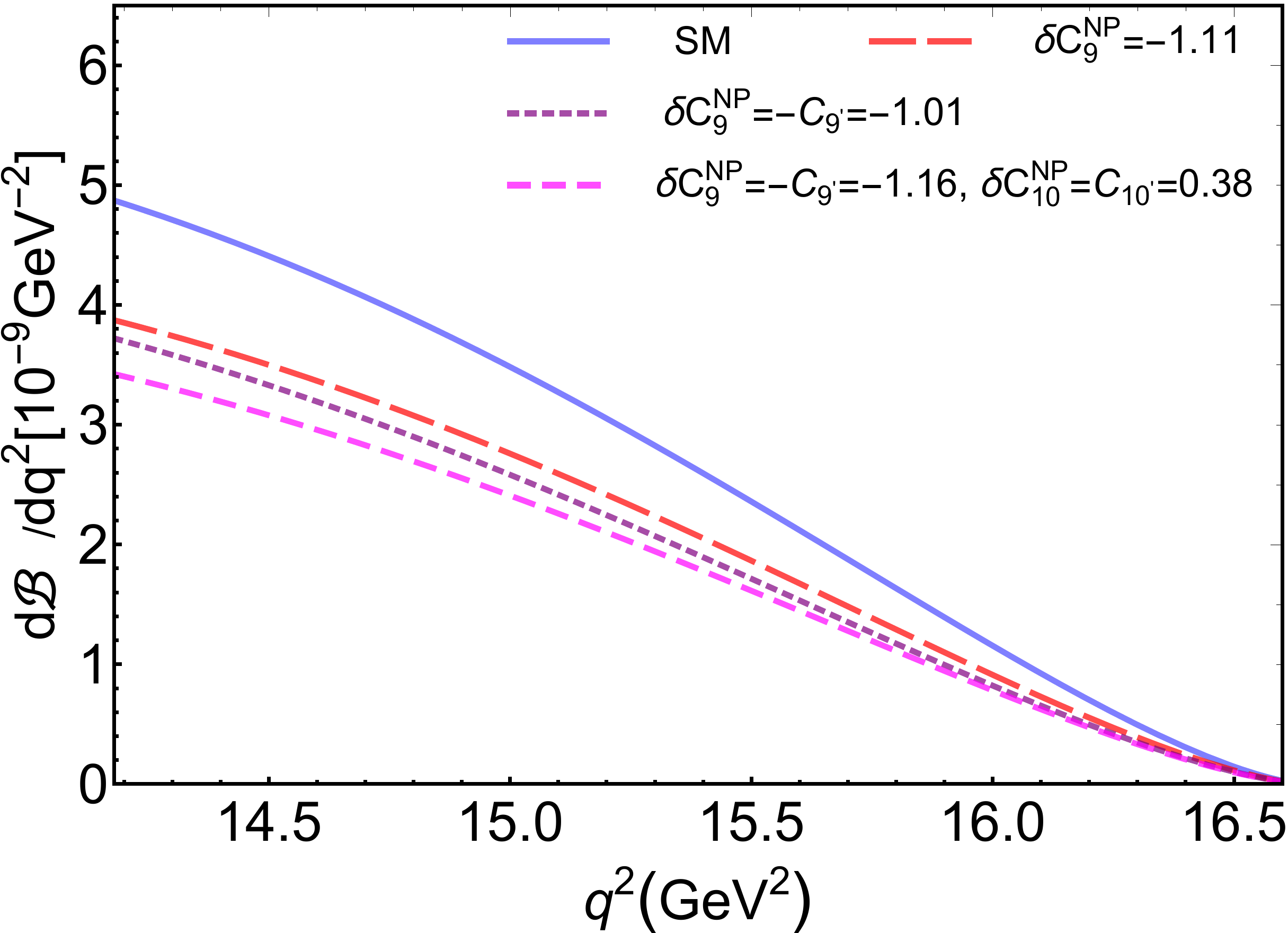}
		\includegraphics[scale=0.3]{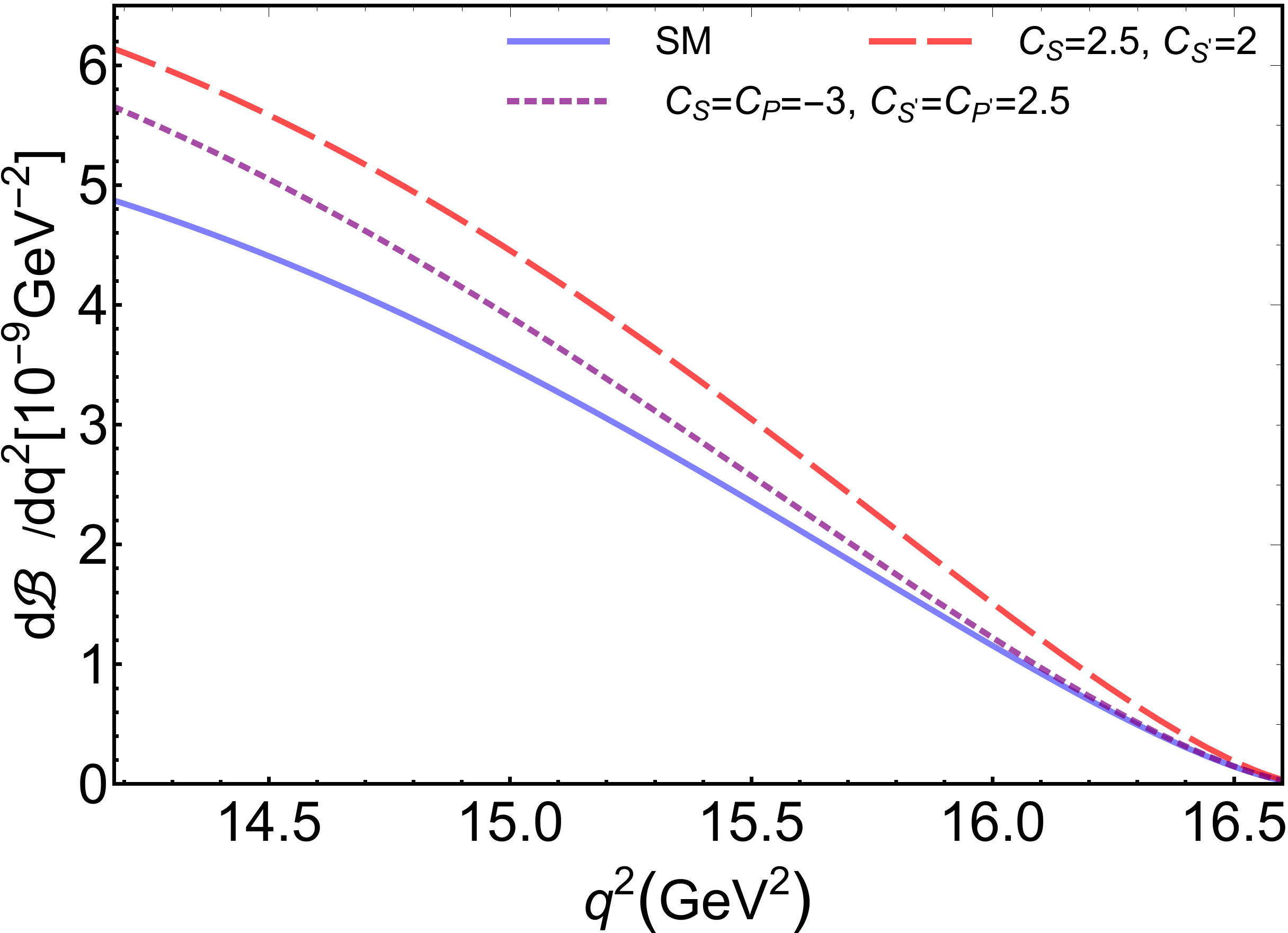}
		\caption{The differential branching ratio of $\Lb\to\Lst(\to N\bar{K})\mu^+\mu^-$ is shown in the SM (blue solid line) and in the NP (dashed colored) at the low-recoil. The figures are for illustrative purpose only (see text for details). \label{fig:NP1}}
	\end{center}
\end{figure}

In the light of the above discussions, our numerical analysis come with the following caveat. The analysis is preliminary and the plots are for illustrative purpose only. Accurate predictions require very careful reassessments of our simplifying assumptions. 

In order to ascertain the effects of the NP, we first discuss the model-independent constraints on the Wilson coefficients. For the simplicity of the discussion, we consider two scenarios-- (i) only the chirality flipped NP (SM+SM$^\prime$) is present, and (ii) only scalar NP is present (SM+SP). For scenario (i) model-independent constraints on $\delta\mC_{9,10}^{\rm NP}$ and  $\mC_{9^\prime, 10^\prime}$ are available from the global fits to $b\to s\mu^+\mu^-$ data. For demonstration we restrict ourselves to the following benchmark solutions \cite{Capdevila:2018jhy,Capdevila:2017bsm, Blake:2019guk} 
\begin{align}
&\delta \mC_{9}^{\rm NP} = -1.11\, ,\\
&\delta \mC_{9}^{\rm NP} = - \mC_{9^\prime} = -1.01\,,\\
& \delta \mC_{9}^{\rm NP} = - \mC_{9^\prime} = -1.16\, ,\quad \delta \mC_{10}^{\rm NP} = \mC_{10^\prime} = 0.38\, .
\end{align}

\begin{figure}[h!]
	\begin{center}
		\includegraphics[scale=0.3]{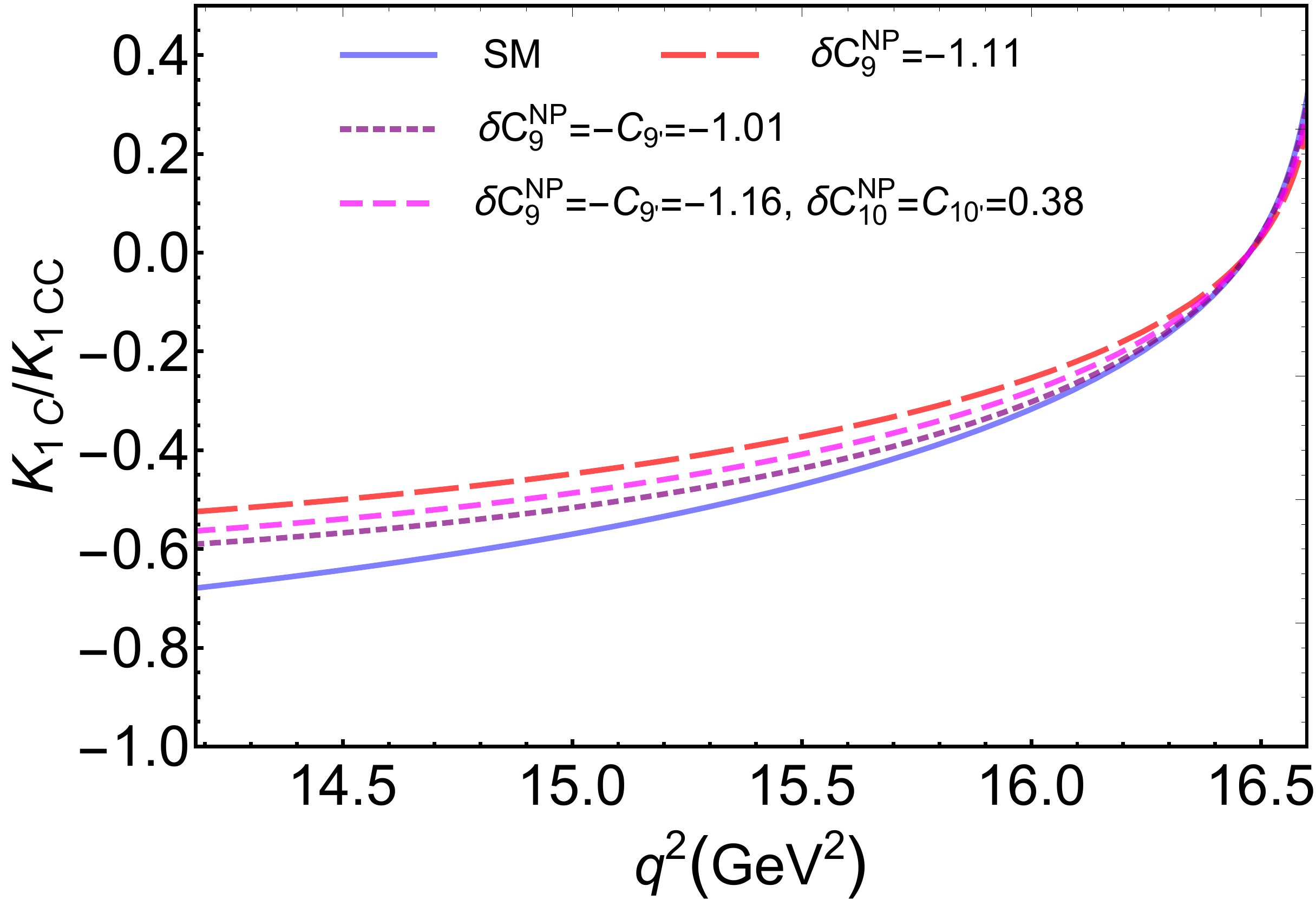}
		\includegraphics[scale=0.3]{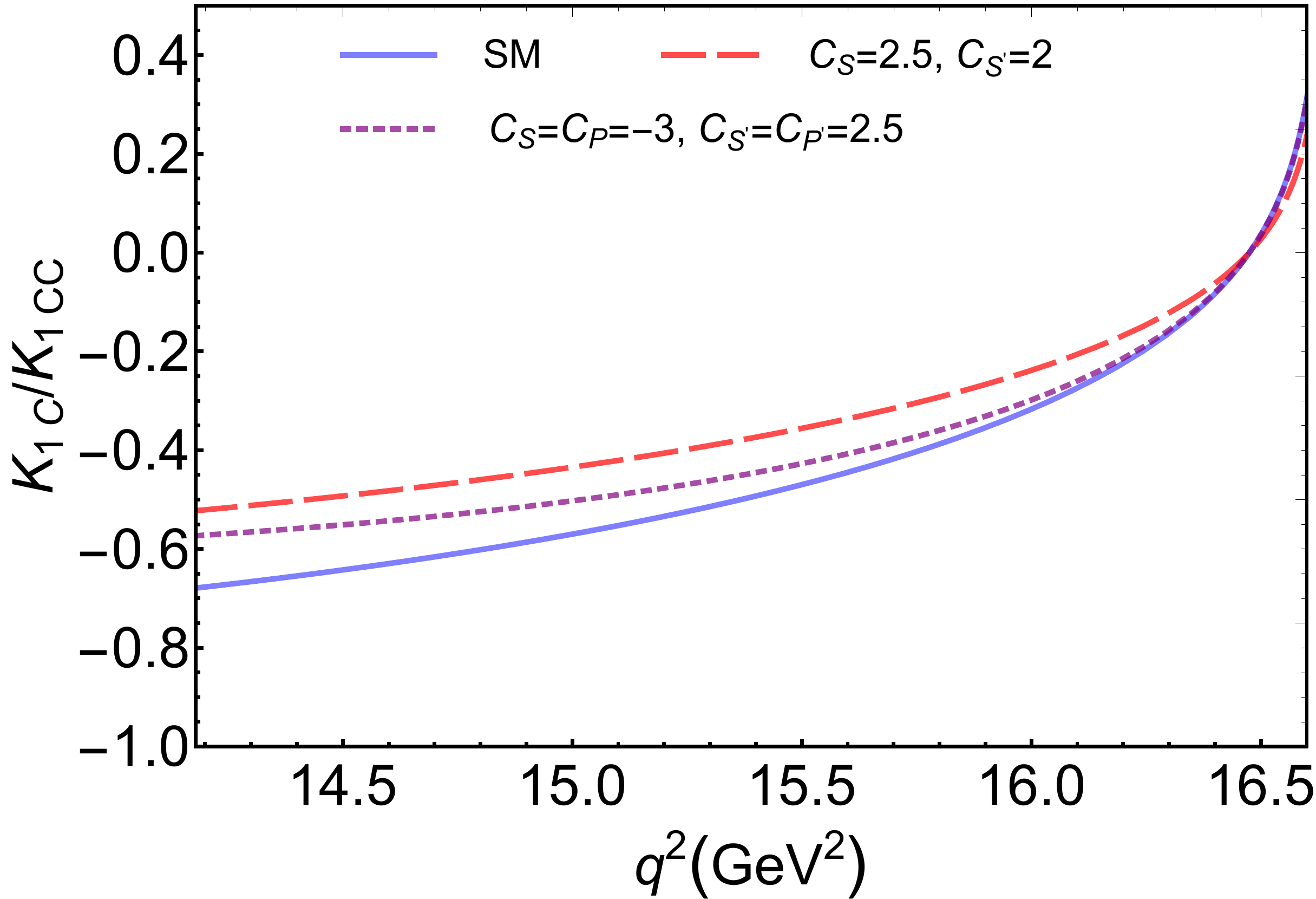}
		\includegraphics[scale=0.3]{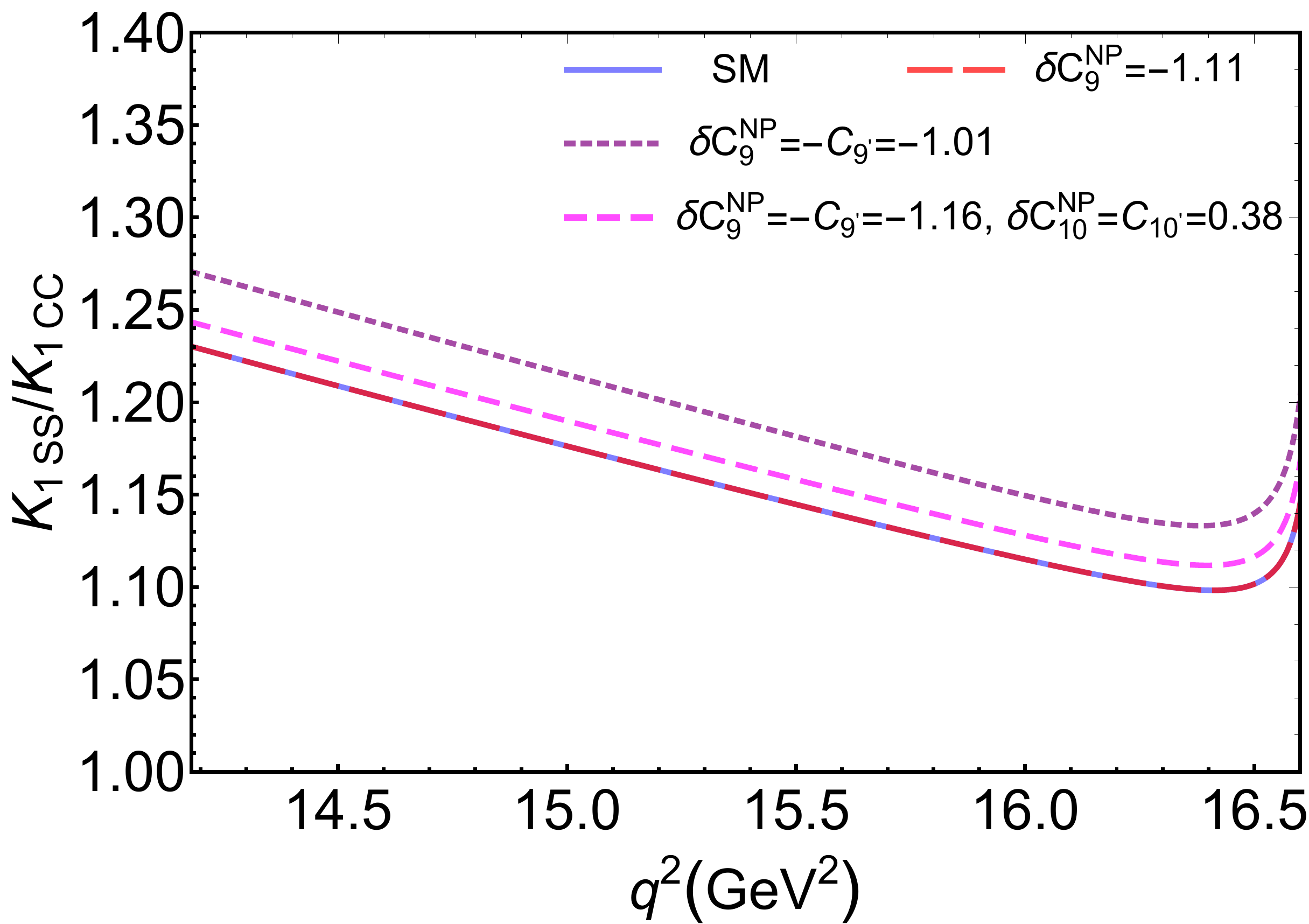}
		\includegraphics[scale=0.3]{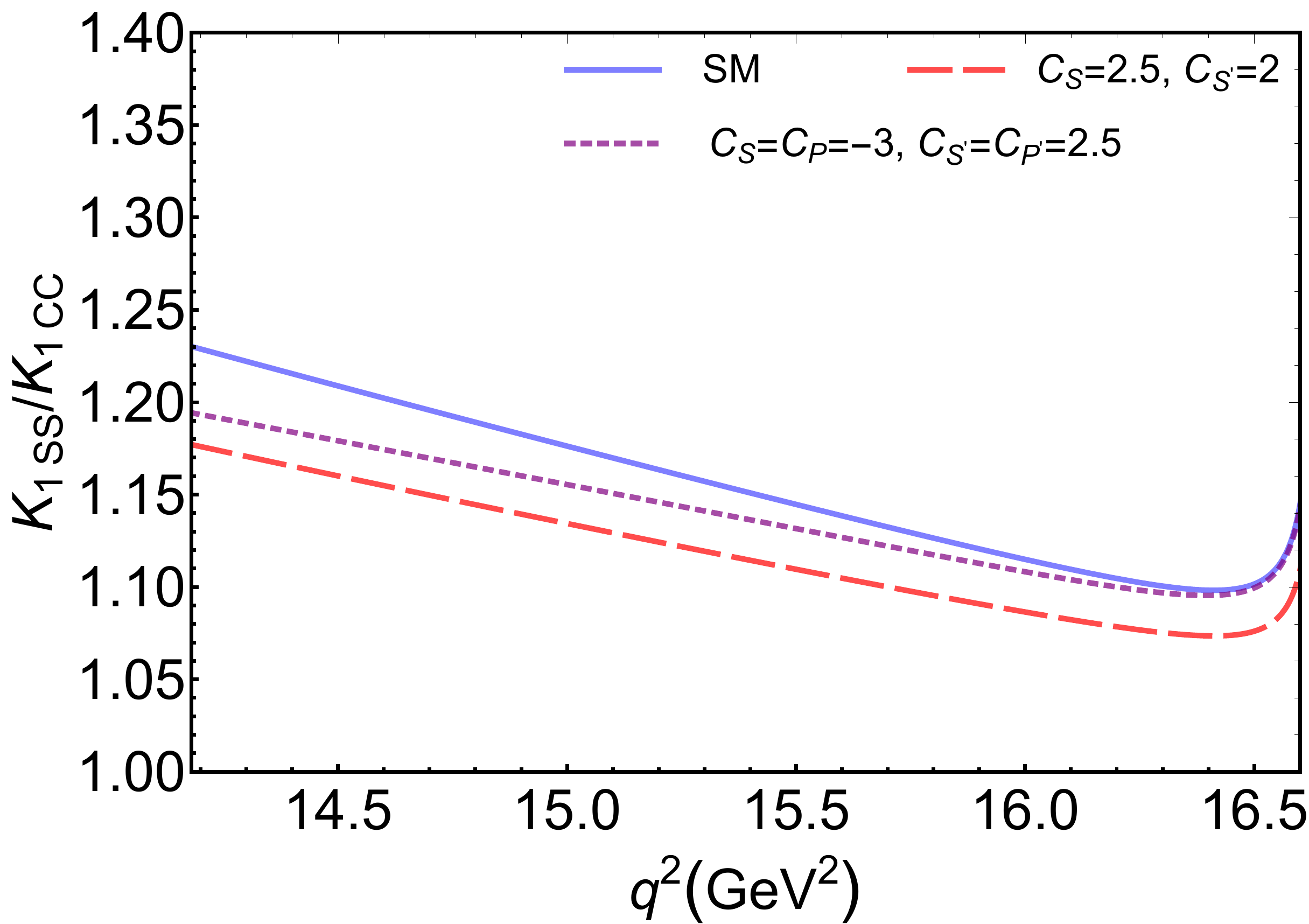}
		\caption{The ratios $K_{1c}/K_{1cc}$ and $K_{1ss}/K_{1cc}$ of $\Lb\to\Lst(\to N\bar{K})\mu^+\mu^-$ in the SM (blue solid line) and in the NP (dashed colored) at the low-recoil. The figures are for illustrative purpose only (see text for details). \label{fig:NP2}}
	\end{center}
\end{figure}

In scenario (ii) model-independent constraints on $\mC_{S^{(\prime)}, P^{(\prime)}}$ come from the experimental data on $B_s\to\mu^+\mu^-$ and the inclusive $\bar{B}\to X_s\mu^+\mu^-$ decays  \cite{Das:2018sms}
\begin{equation}
\mC_{S^{(\prime)}, P^{(\prime)}} = [-4.0, +4.0]\, .
\end{equation}
To the best of our knowledge there is no model-independent global fit that considers a scenario where both chirality flipped and scalar operators are simultaneously present. This scenario is therefore not discussed in this paper.

In Fig.~\ref{fig:SM1} we have shown the SM estimates of the observables $d\mathcal{B}/dq^2$, $A_{\rm FB}^\ell$ and $F_L$. Our choice for the low-recoil phase space is $14.2\le q^2 \le (m_\Lb-m_\Lst)^2$. The bands correspond to different sources of uncertainties. The blue bands correspond to the form factor uncertainties, the red bands correspond to the corrections to the improved Isgur-Wise relations, and the blue bands correspond to the power corrections of the order $\mathcal{O}(\alpha_s\Lambda/m_b)$. Due to the lack of realistic estimate of form factor uncertainties at present, we have ignored the uncertainties of the various inputs given in table \ref{tab:inputs}.  In the low-recoil region the masses of the muon has negligible effect and hence we put $m_\mu=0$.  In Fig.~\ref{fig:RatSMlow} we have shown the SM estimates of the two ratios $K_{1c}/K_{1cc}$ and $K_{1ss}/K_{1cc}$.

In figure \ref{fig:NP1} we have presented the NP sensitivities of the differential branching ratio of $\Lb\to\Lst(\to N\!\bar{K})\mu^+\mu^-$ in the low-recoil region. In figure \ref{fig:NP2} we have shown the NP sensitivities the ratios $K_{1c}/K_{1cc}$ and $K_{1ss}/K_{1cc}$ at low-recoil. In the NP plots, the lines correspond to the central values of all the inputs. To avoid clutter, in all the NP plots we have not shown the errors bands coming from form factors and other sources. With a future determination of the form factors in lattice QCD \cite{Meinel}, our determinations can be improved.

\section{Summary \label{sec:summary}}
In this paper we have studied some aspects of the $\Lb\to\Lst(\to N\!\bar{K})\ell^+\ell^-$ decay. The underlying $b\to s\ell^+\ell^-$ effective Hamiltonian is extended by including the chirality flipped counterparts of the Standard Model operators, and scalar and pseudo-scalar operators. We have presented a full angular analysis where we have also retained the masses of the final state leptons. The angular observables are expressed in terms of the transversity amplitudes. The four-fold distribution allows us to construct several observables that we study in the Standard Model and in model-independent new physics. 

There are fourteen form factors that contribute to the decay. To reduce the uncertainties coming from them, we have exploited the Heavy Quark Effective theory in the low-recoil. Working at the leading order in the $1/m_b$  and including the $\mathcal{O}(\alpha_s)$ corrections we have derived improved Isgur-Wise relations between the form factors. These relations correlate them as a result of which the transversity amplitudes depend on single form factors. By means of these transversity amplitudes, the short- and long-distance physics factorizes in the angular observables. The low-recoil factorization helps us construct observables from which short-distance physics can be tested with minimal form factor inputs. Alternatively, if new physics is not present, the form factor can be tested without interference from short-distance physics. 

In the absence of any lattice QCD calculation, we have taken the form factors from quark model calculations. For our new physics analysis, we have used the model-independent constraints on the new physics Wilson coefficients. Improved predictions in this decay will be possible in the future when the form factors are available from calculations in the lattice QCD.

\section*{Acknowledgements}
The authors would like to thank Debajyoti Choudhury for fruitful discussions and Stefan Meinel for useful communications. We also thank Sebastien Descotes-Genon and Mart\'in Novoa Brunet for useful communications regarding reference \cite{Descotes-Genon:2019dbw}. DD acknowledges the DST, Govt. of India for the INSPIRE Faculty Fellowship (grant number IFA16-PH170). JD acknowledges the Council of Scientific and Industrial Research (CSIR), Govt. of India for JRF fellowship grant with File No. 09/045(1511)/2017-EMR-I. 

\appendix

\section{Decay kinematics \label{app:kin}}
In this section we describe our kinematics following LHCb's convention for $\Lambda_b\to\Lambda(\to N\pi)\ell^+\ell^-$ decay up to the identification $\theta_\Lst=\theta_b$ and $\phi=\chi$ \cite{Aaij:2018gwm,Aaij:2015xza}. The lepton $\ell^-$ has momentum $q_2$ and makes an angle $\theta_\ell$ with respect to the $+z$ axis in the dilepton rest frame (denoted as $2\ell$-RF). Therefore, $q^\mu_{1,2}$ read
\begin{equation}
\begin{split}
& q_1^\mu\big|_{2\ell-\rm RF} = (E_\ell, -|q_{2\ell}|\sin\theta_\ell,0,-|q_{2\ell}|\cos\theta_\ell)\, ,\\
& q_2^\mu\big|_{2\ell-\rm RF} = (E_\ell, |q_{2\ell}|\sin\theta_\ell,0,|q_{2\ell}|\cos\theta_\ell)\, ,
\end{split}
\end{equation}
where 
\begin{equation}
q_{2\ell}=\beta_\ell \frac{\sqrt{q^2}}{2},\quad \beta_\ell = \sqrt{1-\frac{4m_\ell^2}{q^2}}.
\end{equation}
Similarly, in the $N\!\bar{K}$ rest frame (denoted as $N\!\bar{K}$-RF), characterized by $k^2=m_\Lst^2$, the four-momentum $k^\mu_{1,2}$ read
\begin{equation}
\begin{split}
&k_1^\mu\big|_{N\!\bar{K}-\rm RF} = (E_N, |k_{N\!\bar{K}}|\sin\theta_\Lst\cos\phi, |k_{N\!\bar{K}}|\sin\theta_\Lst\sin\phi, |k_{N\!\bar{K}}|\cos\theta_\Lst )\, ,\\
&k_2^\mu\big|_{N\!\bar{K}-\rm RF} = (E_K, -|k_{N\!\bar{K}}|\sin\theta_\Lst\cos\phi, -|k_{N\!\bar{K}}|\sin\theta_\Lst\sin\phi, -|k_{N\!\bar{K}}|\cos\theta_\Lst )\, ,
\end{split}
\end{equation}
where
\begin{equation}
E_N = \frac{k^2+m_N^2-m_{\bar{K}}^2}{2\sqrt{k^2}}\, ,\quad E_K = \frac{k^2+m_{\bar{K}}^2-m_N^2}{2\sqrt{k^2}}\, ,
\end{equation}
and 
\begin{equation}
|k_{N\!\bar{K}}| = \frac{\sqrt{\lambda(k^2, m_N^2, m_{\bar{K}}^2)}}{2\sqrt{k^2}}\, .
\end{equation}

\section{Polarizations of the virtual gauge boson \label{app:pol}}
In the dilepton rest frame, the virtual gauge boson polarization four-vectors are
\begin{equation}
\bar{\epsilon}^\mu(\pm) = \frac{1}{\sqrt{2}}(0, \mp 1, -i, 0)\, ,\quad
\bar{\epsilon}^\mu(0) = (0, 0, 0, 1)\, ,\quad
\bar{\epsilon}^\mu(t) = (1, 0, 0, 0)\, .
\end{equation} 
The vectors satisfy the following orthonormality and completeness relations
\begin{equation}
\bar{\epsilon}^{\ast\mu}(n) \bar{\epsilon}_{\mu}(n^\prime) = g_{nn^\prime}\, ,\quad \sum_{n,n^\prime} \bar{\epsilon}^{\ast\mu}(n) \bar{\epsilon}^{\nu}(n^\prime) g_{nn^\prime} = g^{\mu\nu}\, ,\quad n, n^\prime = t, \pm 1, 0\, ,
\end{equation}
where $g_{nn^\prime} =  diag(+1,-1,-1,-1)$ and our choice of the metric tensor is $g^{\mu\nu}=diag(1,-1,$ $-1,-1)$.
%

\section{Lepton helicity amplitudes \label{app:LepHel}}
The explicit expressions of the lepton helicity amplitudes require us to calculate 
\begin{equation}
\bar{u}_{\ell_1} (1\mp\gamma_5) v_{\ell_2}\, ,\quad \bar{\epsilon}^\mu(\lambda) \bar{u}_{\ell_1} \gamma_\mu (1\mp\gamma_5) v_{\ell_2}\, .
\end{equation}
Following \cite{Haber:1994pe} the explicit expressions of the spinor for the lepton $\ell_1$ are 
\begin{align}
& u_{\ell_1}(\lambda) = 
\begin{pmatrix}
\sqrt{E_\ell+m_\ell} \chi^u_\lambda  \\ 2 \lambda \sqrt{E_\ell-m_\ell} \chi^u_\lambda
\end{pmatrix}\, ,
\quad \chi^u_{+\frac{1}{2}} = \begin{pmatrix} \cos\frac{\theta_\ell}{2} \\ \sin\frac{\theta_\ell}{2} \end{pmatrix}\, \nn\\
&\chi^u_{-\frac{1}{2}} = \begin{pmatrix} -\sin\frac{\theta_\ell}{2} \\ \cos\frac{\theta_\ell}{2} \end{pmatrix}\, .
\end{align}
For the second lepton $\ell_2$ which is moving in the opposite direction to $\ell_1$, the two component spinor $\chi^v$ looks like
\begin{equation}
\chi^v_{-\lambda} = \xi_\lambda \chi^u_{\lambda}\, ,\quad \xi_\lambda = 2\lambda e^{-2i\lambda\phi}\, .
\end{equation}
Hence we have 
\begin{align}
& v_{\ell_2}(\lambda) = 
\begin{pmatrix}
\sqrt{E_\ell-m_\ell} \chi^v_{-\lambda}  \\ -2 \lambda \sqrt{E_\ell+m_\ell} \chi^v_{-\lambda}
\end{pmatrix}\, ,
\quad \chi^v_{+\frac{1}{2}} = \begin{pmatrix} \sin\frac{\theta_\ell}{2} \\ -\cos\frac{\theta_\ell}{2} \end{pmatrix}\, \nn\\
&\chi^v_{-\frac{1}{2}} = \begin{pmatrix} \cos\frac{\theta_\ell}{2} \\ \sin\frac{\theta_\ell}{2} \end{pmatrix}\, .
\end{align}

With these expressions for the spinors we obtain the following expression of the lepton helicity amplitudes
 \begin{align}\label{eq:LepHel}
 & L^{\plpl}_L = -L^{\mimi}_R = \sqrt{q^2}(1+\beta_\ell)\, ,\quad L^{\mimi}_L = -L^{\plpl}_R = \sqrt{q^2}(1-\beta_\ell)\, ,\nn\\
 & L^{\plpl}_{L,-1}=L^{\plpl}_{R,-1}=L^{\mimi}_{L,+1}=L^{\mimi}_{R,+1} = \sqrt{2}m_\ell\sin\theta_\ell\, ,\nn\\
 & L^{\mimi}_{L,-1}=L^{\mimi}_{R,-1}=L^{\plpl}_{L,+1}=L^{\plpl}_{R,+1} = -\sqrt{2}m_\ell\sin\theta_\ell\, ,\nn\\
 & L^{\plmi}_{L,-1} = - L^{\mipl}_{R,+1} = -\sqrt{\frac{q^2}{2}}(1-\beta_\ell)(1-\cos\theta_\ell)\, ,\nn\\
 & L^{\mipl}_{L,-1} = -L^{\plmi}_{R,+1} = \sqrt{\frac{q^2}{2}}(1+\beta_\ell)(1+\cos\theta_\ell)\, ,\\
 & L^{\plmi}_{R,-1}=-L^{\mipl}_{L,+1} = -\sqrt{\frac{q^2}{2}}(1+\beta_\ell)(1-\cos\theta_\ell)\, ,\nn\\
 & L^{\mipl}_{R,-1}=-L^{\plmi}_{L,+1} = \sqrt{\frac{q^2}{2}}(1-\beta_\ell)(1+\cos\theta_\ell)\, ,\nn\\
 & L^{\plpl}_{L,0}=-L^{\mimi}_{L,0}=+L^{\plpl}_{R,0}=-L^{\mimi}_{R,0} = 2m_\ell\cos\theta_\ell\, ,\nn\\
 & L^{\plmi}_{L,0}=L^{\mipl}_{R,0} = -\sqrt{q^2}(1-\beta_\ell)\sin\theta_\ell\, ,\quad L^{\mipl}_{L,0} = L^{\plmi}_{R,0} = -\sqrt{q^2}(1+\beta_\ell)\sin\theta_\ell\, ,\nn\\
 & L^{\plpl}_{L,t} = L^{\mimi}_{L,t} = -L^{\plpl}_{R,t} = -L^{\mimi}_{R,t} = 2m_\ell\, .\nn
 \end{align}
 %

\section{$\Lambda_b\to\Lambda^\ast$ matrix elements \label{app:ffParam}}
The $\Lambda_b\to\Lambda^\ast$ transition matrix element can be parametrized using the $\Lambda_b$ and $\Lambda^\ast$ spinors in terms of form factors. For the transitions through vector, and axial vector  currents, the matrix elements are parametrized in terms of eight form factors \cite{Descotes-Genon:2019dbw}
\begin{align}\label{eq:ffVA1}
\langle\Lst | \bar s \gamma^\mu b|\Lb\rangle=
& \bar u_\alpha(k,s_\Lst)\biggl\{p^\alpha\biggl[f_t^V(q^2) (m_{\Lb}-m_\Lst)\frac{q^\mu}{q^2}\nn\\
&+ f_0^V(q^2) \frac{m_{\Lb}+m_\Lst}{s_+}(p^\mu +k^\mu-\frac{q^\mu}{q^2}(m_{\Lb}^2-m_\Lst^2))\nn\\
&+ f_\perp^V(q^2)(\gamma^\mu-2\frac{m_\Lst}{s_+}p^\mu -2\frac{m_{\Lb}}{s_+}k^\mu)\biggr]\nn\\
&+ f_g^V(q^2) \left[g^{\alpha\mu}+m_\Lst\frac{p^\alpha}{s_-} \left(\gamma^\mu - 2 \frac{k^\mu}{m_\Lst} +2 \frac{m_\Lst p^\mu +m_\Lb k^\mu}{s_+}\right)\right]\biggr\}u(p,s_{\Lb})\,,\\
\langle\Lst | \bar s \gamma^\mu \gamma ^5 b|\Lb\rangle=& -\bar u_\alpha(k,s_\Lst)\gamma^5\biggl\{p^\alpha\biggl[
f_t^A(q^2) (m_{\Lb}+m_\Lst)\frac{q^\mu}{q^2}\nn\\
&+ f_0^A(q^2) \frac{m_{\Lb}-m_\Lst}{s_-}(p^\mu +k^\mu-\frac{q^\mu}{q^2}(m_{\Lb}^2-m_\Lst^2))\nn\\
&+ f_\perp^A(q^2)(\gamma^\mu+2\frac{m_\Lst}{s_-}p^\mu -2\frac{m_{\Lb}}{s_-}k^\mu)\biggr]\nn\\
&+ f_g^A(q^2) \left[g^{\alpha\mu}-m_\Lst\frac{p^\alpha}{s_+} \left(\gamma^\mu + 2 \frac{k^\mu}{m_\Lst} -2 \frac{m_\Lst p^\mu -m_\Lb k^\mu}{s_-}\right)\right]\biggr\}u(p,s_{\Lb})\,.
\end{align}
Applying the equation of motion on \eqref{eq:ffVA1} we get the matrix elements for the $\bar{s}b$ and $\bar{s}\gamma_5b$ as
\begin{align}
& \langle\Lst | \bar s b|\Lb\rangle = f^V_t \frac{\mLb-\mL}{m_b-m_s} u_\alpha(k,s_\Lst) p^\alpha u(p,s_{\Lb})\, ,  \\
&\langle\Lst | \bar s \gamma^\mu \gamma ^5 b|\Lb\rangle = f^A_t \frac{\mLb+\mL}{m_b+m_s} u_\alpha(k,s_\Lst) p^\alpha \gamma_5 u(p,s_{\Lb})\,.
\end{align}

The matrix elements corresponding to the tensor and axial-tensor currents are parametrized in terms of six form factors
%
%
\begin{align}
\langle\Lst | \bar si \sigma^{\mu\nu}q_\nu b|\Lb\rangle=& - \bar u_\alpha(k,s_\Lst)\biggl\{p^\alpha\biggl[f_0^T(q^2) \frac{q^2}{s_+}(p^\mu +k^\mu-\frac{q^\mu}{q^2}(m_{\Lb}^2-m_\Lst^2))\nn\\
&+ f_\perp^T(q^2)(m_{\Lb} + m_\Lst)(\gamma^\mu-2\frac{m_\Lst}{s_+}p^\mu -2\frac{m_{\Lb}}{s_+}k^\mu)\biggr]\nn\\
&+ f_g^T(q^2)\left[g^{\alpha\mu}+m_\Lst\frac{p^\alpha}{s_-} \left(\gamma^\mu - 2 \frac{k^\mu}{m_\Lst} +2 \frac{m_\Lst p^\mu +m_\Lb k^\mu}{s_+}\right)\right]\biggr\}u(p,s_{\Lb}),\\
\label{eq:ffT}
\langle\Lst | \bar si \sigma^{\mu\nu}\gamma^5q_\nu b|\Lb\rangle=& - \bar u_\alpha(k,s_\Lst)\gamma^5\biggl\{p^\alpha\biggl[f_0^{T5}(q^2) \frac{q^2}{s_-}(p^\mu +k^\mu-\frac{q^\mu}{q^2}(m_{\Lb}^2-m_\Lst^2))\nn\\
&+ f_\perp^{T5}(q^2)(m_{\Lb} - m_\Lst)(\gamma^\mu+2\frac{m_\Lst}{s_-}p^\mu -2\frac{m_{\Lb}}{s_-}k^\mu)\biggr]\nn\\
&+ f_g^{T5}(q^2)\left[g^{\alpha\mu}-m_\Lst\frac{p^\alpha}{s_+} \left(\gamma^\mu + 2 \frac{k^\mu}{m_\Lst} -2 \frac{m_\Lst p^\mu -m_\Lb k^\mu}{s_-}\right)\right]\biggr\}u(p,s_{\Lb}).
\end{align}

\section{Helicity to Transversity \label{app:TAs}}
Using the definitions of the helicity amplitudes \eqref{eq:Hl1}-\eqref{eq:Hl2}, we construct the transversity amplitudes for the (axial-)vector currents
\begin{align}
\begin{split}
& B_{\perp_1}^{L(R)} = \frac{N}{\sqrt{2}}\bigg[H^{L(R)}_{\rm VA,+}(-1/2,-3/2) + H^{L(R)}_{\rm VA,-}(+1/2,+3/2)  \bigg]\, ,\\
& B_{\|_1}^{L(R)} = \frac{N}{\sqrt{2}}\bigg[H^{L(R)}_{\rm VA,+}(-1/2,-3/2) - H^{L(R)}_{\rm VA,-}(+1/2,+3/2)  \bigg]\, ,\\
& A_{\perp_0}^{L(R)} = \frac{N}{\sqrt{2}}\bigg[H^{L(R)}_{\rm VA,0}(+1/2,+1/2) + H^{L(R)}_{\rm VA,0}(-1/2,-1/2)  \bigg]\, ,\\
& A_{\|_0}^{L(R)} = \frac{N}{\sqrt{2}}\bigg[H^{L(R)}_{\rm VA,0}(+1/2,+1/2) - H^{L(R)}_{\rm VA,0}(-1/2,-1/2)  \bigg]\, ,\\
& A_{\perp_1}^{L(R)} = \frac{N}{\sqrt{2}} \bigg[H^{L(R)}_{\rm VA,+}(+1/2,-1/2) + H^{L(R)}_{\rm VA,-}(-1/2,+1/2) \bigg]\, ,\\
& A_{\|_1}^{L(R)} = \frac{N}{\sqrt{2}} \bigg[H^{L(R)}_{\rm VA,+}(+1/2,-1/2) - H^{L(R)}_{\rm VA,-}(-1/2,+1/2) \bigg]\, ,\\
& A_{\perp_t}^{L(R)} = \frac{N}{\sqrt{2}} \bigg[ H_{{\rm VA},t}^{L(R)}(+1/2,+1/2) + H_{{\rm VA},t}^{L(R)}(-1/2,-1/2) \bigg]\, ,\\
\label{eq:TAVA2}
& A_{\|_t}^{L(R)} = \frac{N}{\sqrt{2}} \bigg[ H_{{\rm VA},t}^{L(R)}(+1/2,+1/2) - H_{{\rm VA},t}^{L(R)}(-1/2,-1/2) \bigg]\, ,
\end{split}
\end{align} 
and for the scalar and pseudo-scalar operators
\begin{align}
\begin{split}
A_{\perp S}^{L(R)} = \frac{N}{\sqrt{2}} \bigg[ H_{\rm SP}^{L(R)}(+1/2,+1/2) + H_{\rm SP}^{L(R)}(-1/2,-1/2)   \bigg]\, ,\\
A_{\| S}^{L(R)} = \frac{N}{\sqrt{2}} \bigg[ H_{\rm SP}^{L(R)}(+1/2,+1/2) - H_{\rm SP}^{L(R)}(-1/2,-1/2)   \bigg]\, .
\end{split}
\end{align}

\section{The Rarita-Schwinger spinor solutions \label{app:Rarita}}
The solutions of Rarita-Schwinger spinors in the $\Lambda^\ast$ rest frame are \cite{Huang:2003ym} 
\begin{equation}
\begin{aligned}
(U^{-3/2})^\mu=&\sqrt{m_\Lst}\left(\begin{array}{cccc} 
0 & 0 & 0 & 0\\
0 & 1 & 0 & 0\\
0 & -i & 0 & 0\\
0 & 0 & 0 & 0
\end{array}\right),\qquad
(U^{-1/2})^\mu=\sqrt{\frac{m_\Lst}{3}}\left(\begin{array}{cccc} 
0 & 0 & 0 & 0\\
1 & 0 & 0 & 0\\
-i & 0 & 0 & 0\\
0 & 2 & 0 & 0
\end{array}\right),\\
(U^{+1/2})^\mu=&\sqrt{\frac{m_\Lst}{3}}\left(\begin{array}{cccc} 
0 & 0 & 0 & 0\\
0 & -1 & 0 & 0\\
0 & -i & 0 & 0\\
2 & 0 & 0 & 0
\end{array}\right),\qquad
(U^{+3/2})^\mu=\sqrt{m_\Lst}\left(\begin{array}{cccc} 
0 & 0 & 0 & 0\\
-1 & 0 & 0 & 0\\
-i & 0 & 0 & 0\\
0 & 0 & 0 & 0
\end{array}\right).
\end{aligned}
\end{equation}
The solutions of the Dirac spinors corresponding to the neucleon are \cite{Haber:1994pe}
\begin{equation}
u^{+1/2}=\frac{1}{2m_\Lst}\left(\begin{array}{c} 
\sqrt{r_+}\cos\frac{\theta_\Lst}{2}\\ 
\sqrt{r_+}\sin\frac{\theta_\Lst}{2}e^{i\phi}\\ 
\sqrt{r_-}\cos\frac{\theta_\Lst}{2}\\
\sqrt{r_-}\sin\frac{\theta_\Lst}{2}e^{i\phi} \end{array}\right), \qquad
u^{-1/2}=\frac{1}{2m_\Lst}\left(\begin{array}{c} 
-\sqrt{r_+}\sin\frac{\theta_\Lst}{2}e^{-i\phi}\\ 
\sqrt{r_+}\cos\frac{\theta_\Lst}{2}\\ 
\sqrt{r_-}\sin\frac{\theta_\Lst}{2}e^{-i\phi} \\
-\sqrt{r_-}\cos\frac{\theta_\Lst}{2}  \end{array}\right) .
\end{equation}

\section{Angular Coefficients \label{app:angular}}
The expression of the angular coefficients are 
\begin{align}
& \mathcal{K}_{1c}=-2\beta_\ell\bigg(\re(A_{\perp1}^L A_{\parallel1}^{L\ast})-\{L\leftrightarrow R\} \bigg)\, ,\nn\\\
&\mathcal{K}_{1c}^\prime=-2\beta_\ell\bigg(\re(\apasl\apaol^{\ast})+\re(\apasr\apaol^{\ast})+\re(\apesl\apeol^{\ast})+\re(\apesr\apeol^{\ast}) + \{L\leftrightarrow R\} \bigg)\, ,\nn\\
& \mathcal{K}_{1c}^{\prime\prime}=0\, ,\\\
&\mathcal{K}_{1cc}
=\bigg(|\apanl|^2+|\apasl|^2+|\apenl|^2+|\apesl|^2+ \{L\leftrightarrow R\} \bigg)\, ,\nn\\\
&\mathcal{K}_{1cc}^\prime=2\bigg(-\re(\apatr\apasl^{\ast})+\re(\apasl\apatl^{\ast})-\re(\apetr\apesl^{\ast})+\re(\apesl\apetl^{\ast}) + \{ L\leftrightarrow R\} \bigg)\,,\nn\\\
&\mathcal{K}_{1cc}^{\prime\prime} = 2 \bigg(|\apaol|^2 -|\apanl|^2-|\apasl|^2+|\apatl|^2+|\apeol|^2-|\apenl|^2-|\apesl|^2+|\apetl|^2\nn\\\
&\quad\quad+\re(\apaor\apaol^{\ast})+\re(\apanr\apanl^{\ast})-\re(\apasr\apasl^{\ast})-\re(\apatr\apatl^{\ast})\nn\\\
&\quad\quad+\re(\apeor\apeol^{\ast})+\re(\apenr\apenl^{\ast})-\re(\apesr\apesl^{\ast})-\re(\apetr\apetl^{\ast})\nn\\\
&\quad\quad+ \{ L\leftrightarrow R\} \bigg)\, ,\\
&\mathcal{K}_{1ss}=\frac{1}{2}\bigg(2|\apaol|^2+|\apanl|^2+2|\apasl|^2+2|\apeol|^2+|\apenl|^2+2|\apesl|^2 + \{ L\leftrightarrow R \}\bigg) \nn\\
&\mathcal{K}_{1ss}^\prime=-2\bigg(\re(\apatr\apasl^{\ast})-\re(\apasl\apatl^{\ast})+\re(\apetr\apesl^{\ast})-\re(\apesl\apetl^{\ast}) + \{ L\leftrightarrow R \}\bigg)\, ,\nn\\\
&\mathcal{K}_{1ss}^{\prime\prime}= 2 \bigg(-|\apaol|^2-|\apasl|^2+|\apatl|^2-|\apeol|^2-|\apesl|^2+|\apetl|^2\nn\\\
&\quad\quad+\re(\apaor\apaol^{\ast})+\re(\apanr\apanl^{\ast})-\re(\apasr\apasl^{\ast})-\re(\apatr\apatl^{\ast}) +\re(\apeor\apeol^{\ast}) \nn\\\
&\quad\quad+\re(\apenr\apenl^{\ast})-\re(\apesr\apesl^{\ast})-\re(\apetr\apetl^{\ast}) + \{ L\leftrightarrow R \}\bigg)\, ,\\
&\mathcal{K}_{2c}=-\frac{1}{2}\beta_\ell\bigg(\re(\apenl\apanl^{\ast})+3\re(\bpenl\bpanl^{\ast})- \{ L\leftrightarrow R \}\bigg)\, ,\nn\\\
&\mathcal{K}_{2c}^{\prime}=-\frac{1}{2}\beta_\ell\bigg(\re(\apasl\apaol^{\ast})+\re(\apasl\apaor^{\ast})+\re(\apesl\apeol^{\ast})+\re(\apesl\apeor^{\ast}) + \{ L\leftrightarrow R \}\bigg)\, ,\nn\\\
&\mathcal{K}_{2c}^{\prime\prime}=0\, ,\\
&\mathcal{K}_{2cc}=\frac{1}{4}\bigg(|\apanl|^2+|\apasl|^2+3|\bpanl|^2+|\apenl|^2+|\apesl|^2+3|\bpenl|^2 + \{ L\leftrightarrow R \}\bigg)\, ,\nn\\\
&\mathcal{K}_{2cc}^\prime=-\frac{1}{2}\bigg(\re(\apatr\apasl^{\ast})-\re(\apasl\apatl^{\ast})+\re(\apetr\apesl^{\ast})-\re(\apesl\apetl^{\ast})+ \{ L\leftrightarrow R \}\bigg)\, ,\nn\\\
&\mathcal{K}_{2cc}^{\prime\prime}=\frac{1}{2}\bigg(|\apaol|^2-|\apanl|^2-|\apasl|^2+|\apatl|^2+|\apeol|^2-|\apenl|^2-|\apesl|^2+|\apetl|^2\nn\\\
&\quad\quad-3|\bpanl|^2-3|\bpenl|^2+\re(\apaor\apaol^{\ast})+\re(\apanr\apanl^{\ast})-\re(\apasr\apasl^{\ast})-\re(\apatr\apatl^{\ast})\nn\\\
&\quad\quad+\re(\apeor\apeol^{\ast})+\re(\apenr\apenl^{\ast})-\re(\apesr\apesl^{\ast})-\re(\apetr\apetl^{\ast})\nn\\\
&\quad\quad+3\re(\bpanr\bpanl^{\ast})+3\re(\bpenr\bpenl^{\ast})+ \{L\leftrightarrow R\}\bigg)\, ,\\
&\mathcal{K}_{2ss}=\frac{1}{8}\bigg(2|\apaol|^2+|\apanl|^2+2|\apasl|^2+2|\apeol|^2+|\apenl|^2+2|\apesl|^2\nn\\
&\quad\quad+3|\bpanl|^2+3|\bpenl|^2-2\sqrt{3}\re(\bpanl\apanl^{\ast})+2\sqrt{3}\re(\bpenl\apenl^{\ast}) + \{L\leftrightarrow R\}\bigg)\, ,\nn\\\
&K_{2ss}^\prime=-\frac{1}{2}\bigg(\re(\apatr\apasl^{\ast})-\re(\apasl\apatl^{\ast})+\re(\apetr\apesl^{\ast})-\re(\apesl\apetl^{\ast})+\{L\leftrightarrow R\}\bigg)\, ,\nn\\\
&\mathcal{K}_{2ss}^{\prime\prime}=\frac{1}{2}\bigg(-|\apaol|^2-|\apasl|^2+|\apatl|^2-|\apeol|^2-|\apesl|^2+|\apetl|^2\nn\\\
&\quad\quad+\re(\apaor\apaol^{\ast})+\re(\apanr\apanl^{\ast})+2\sqrt{3} \re(\bpanl\apanl^{\ast})-\re(\apasr\apasl^{\ast})\nn\\\
&\quad\quad-\re(\apatr\apatl^{\ast})+\re(\apeor\apeol^{\ast})+\re(\apenr\apenl^{\ast})-2\sqrt{3}\re(\bpenl\apenl^{\ast})-\re(\apesr\apesl^{\ast})\nn\\
&\quad\quad-\re(\apetr\apetl^{\ast})
+ 3\re(\bpanr\bpanl^{\ast})+3\re(\bpenr\bpenl^{\ast})+ \{L\leftrightarrow R \}\bigg)\, ,\\
&\mathcal{K}_{3ss}=\frac{\sqrt{3}}{2}\bigg(\re(\bpanl\apanl^{\ast})-\re(\bpenl\apenl^{\ast})+\{L\leftrightarrow R\}\bigg)\, ,\nn\\\
&\mathcal{K}_{3ss}^{\prime}=0\, ,\nn\\\
&\mathcal{K}_{3ss}^{\prime\prime}=-2\sqrt{3}\bigg(\re(\bpanl\apanl^{\ast})-\re(\bpenl\apenl^{\ast})+\{L\leftrightarrow R\}\bigg)\, ,\\
&\mathcal{K}_{4ss}=\frac{\sqrt{3}}{2}\bigg(\im(\bpenl\apanl^{\ast})-\im(\bpanl\apenl^{\ast})+\{L\leftrightarrow R\}\bigg)\, ,\nn\\\
&\mathcal{K}_{4ss}^{\prime}=0\, ,\nn\\\
&\mathcal{K}_{4ss}^{\prime\prime}=-2\sqrt{3}\bigg(\im(\bpenl\apanl^{\ast})-\im(\bpanl\apenl^{\ast})+\{L\leftrightarrow R\}\bigg)\, ,\\
&\mathcal{K}_{5s}=\frac{\sqrt{6}}{2}\beta_\ell\bigg(\re(\bpenl\apaol^{\ast})-\re(\bpanl\apeol^{\ast})-\{L \leftrightarrow R\}\bigg)\, ,\nn\\\
&\mathcal{K}_{5s}^{\prime}= -\frac{\sqrt{6}}{2}\beta_\ell\bigg(\re(\bpanr\apasl^{\ast})-\re(\bpenr\apesl^{\ast})
+\re(\apasl\bpanl^{\ast})-\re(\apesl\bpenl^{\ast})
+\{ L\leftrightarrow R\}\bigg)\, ,\nn\\\
&\mathcal{K}_{5s}^{\prime\prime}=0\, ,\\
&\mathcal{K}_{5sc}=-\frac{\sqrt{6}}{2}\bigg(\re(\bpanl\apaol^{\ast})-\re(\bpenl\apeol^{\ast})+\{L \leftrightarrow R\}\bigg)\, ,\nn\\\
&\mathcal{K}_{5sc}^{\prime}=0\, ,\\
&\mathcal{K}_{5sc}^{\prime\prime}=2\sqrt{6}\bigg(\re(\bpanl\apaol^{\ast})-\re(\bpenl\apeol^{\ast})+\{L \leftrightarrow R\}\bigg)\, ,\nn\\\
&\mathcal{K}_{6s}=\frac{\sqrt{6}}{2}\beta_\ell\bigg(\im(\bpanl\apaol^{\ast})-\im(\bpenl\apeol^{\ast})-\{ L\leftrightarrow R\}\bigg)\, ,\nn\\\
&\mathcal{K}_{6s}^{\prime}= -\frac{\sqrt{6}}{2}\beta_\ell\bigg(\im(\bpenr\apasl^{\ast})-\im(\bpanr\apesl^{\ast})+\im(\apesl\bpanl^{\ast})-\im(\apasl\bpenl^{\ast})
+\{ L\leftrightarrow R\}\bigg)\, ,\\\
&\mathcal{K}_{6s}^{\prime\prime}=0\, ,\nn\\
&\mathcal{K}_{6sc}=-\frac{\sqrt{6}}{2}\bigg(\im(\bpenl\apaol^{\ast})-\im(\bpanl\apeol^{\ast})+\{L\leftrightarrow R\}\bigg)\, ,\nn\\\
&\mathcal{K}_{6sc}^{\prime}=0\, ,\nn\\\
&\mathcal{K}_{6sc}^{\prime\prime}=2\sqrt{6}\bigg(\im(\bpenl\apaol^{\ast})-\im(\bpanl\apeol^{\ast})+\{L \leftrightarrow R\}\bigg)\, .
\end{align}
We agree with the SM results with zero lepton mass presented in Ref.~\cite{Descotes-Genon:2019dbw}.

\section{Numerical inputs \label{sec:inputs}}
The inputs used in our analysis are given in the table below

\begin{longtable}{cr|cr}
	\hline\hline
	inputs  & values  & inputs  & values   \\ 
	\hline 
	$ \alpha_e(m_Z) $  & $1/127.925(16)$ \cite{Patrignani:2016xqp} & $|V_{tb}V_{ts}^\ast|$  & $0.0401 \pm 0.0010$ \cite{Bona:2006ah}  \\ 
	$\mathcal{B}_\Lst$ & $0.45\pm0.01$ \cite{Patrignani:2016xqp} & $\mLb$ & 5.619 GeV \cite{Patrignani:2016xqp} \\ 
	$\mu$ & $4.8$ GeV \cite{Altmannshofer:2008dz} & $\mL$ & $1.5195$ GeV \cite{Patrignani:2016xqp} \\
	$m_b(\overline{\text{MS}})$ & $4.2$ GeV \cite{Altmannshofer:2008dz} & $\tau_{\Lambda_b}$ & $(1.470\pm 0.010)\times 10^{-12} s$ \cite{Patrignani:2016xqp} \\    
	$\alpha_s(M_Z)$ & $0.1181\pm 0.0011$ \cite{Patrignani:2016xqp} & $\mathcal{B}_\Lst$ & $0.45\pm0.01$ \cite{Patrignani:2016xqp}\\
	\hline\caption{List of inputs and their values. \label{tab:inputs}}
\end{longtable}

\end{document}